\documentclass[aps,pra,twocolumn,showpacs]{revtex4}
\usepackage{graphics,pstricks,amsmath}
\usepackage{graphicx}
\usepackage{epsfig}

\begin{document}

\author{Piotr S. \.Zuchowski}
\email{E-mail: piotr.zuchowski@durham.ac.uk} \affiliation{Department
of Chemistry, Durham University, South Road, DH1 3LE, United
Kingdom}

\author{Jeremy M. Hutson}
\email{E-mail: j.m.hutson@durham.ac.uk} \affiliation{Department of
Chemistry, Durham University, South Road, DH1 3LE, United Kingdom}

\title{Low-energy collisions of NH$_3$ and ND$_3$ with
ultracold Rb atoms}

\begin{abstract}
We carry out quantum inelastic scattering calculations of
collisions of Rb atoms with inverting NH$_3$ and ND$_3$
molecules in the energy range between 0 and 100 cm$^{-1}$,
which are important for experiments using velocity-controlled
molecular beams to probe scattering resonances. We focus on
molecules initially in the upper level of the ammonia inversion
doublet for $j=1,k=1$, which is low-field-seeking and can be
controlled in a Stark decelerator. We calculate the integral
elastic and state-to-state inelastic cross sections in the
coupled states approximation. We demonstrate the presence of
both shape and Feshbach resonances in the elastic and inelastic
cross sections at low collision energies and discuss their
origin in terms of the bound states of Rb--ND$_3$ complex. We
also consider elastic and inelastic cross sections in the
ultracold regime, using close-coupling calculations, in order
to assess the viability of sympathetic cooling of ND$_3$ by Rb.
The inelastic cross section for relaxation to the lower level
of the inversion doublet is smaller than expected for such a
strongly coupled system, but is still likely to be too large to
allow sympathetic cooling for ND$_3$ in low-field-seeking
states. However, there is a good prospect that sympathetic
cooling will be possible for molecules in high-field-seeking
states, even when the collision partner is a magnetically
trapped atom in a low-field-seeking state.
\end{abstract}
\date{\today}
\maketitle

\section{Introduction}

Over the last few years, several methods have been developed to
cool molecules to temperatures below 1 degree Kelvin. These
include buffer-gas cooling in cryogenic helium
\cite{Weinstein:CaH:1998, Egorov:2004}, Stark deceleration with
switched electric fields \cite{Bethlem:IRPC:2003,
Bethlem:2006}, velocity filtering \cite{Junglen:2004}, optical
deceleration with laser fields \cite{Fulton:2006}, and crossed
molecular beam scattering \cite{Elioff:2003}. The availability
of cold molecules opens up a new field of low-energy collision
studies, in a novel regime where collisions are dominated by
long-range forces and resonances.

Stark decelerators can be used to {\em control} beam velocity
as well as to reduce it \cite{Heiner:2006}. Pulsed molecular
beams can be generated with much smaller velocity spreads than
is possible with conventional supersonic sources. This offers
the opportunity to study scattering resonances at much higher
resolution than has been possible in the past. Gilijamse {\em
et al.} \cite{Gilijamse:2006} have carried out a
proof-of-concept experiment in which a velocity-controlled beam
of OH radicals collided with Xe atoms in a jet. In this case
the energy resolution of the experiment was limited to 13
cm$^{-1}$ by the velocity spread of the Xe atoms. However,
experiments are under way to collide two velocity-controlled
beams, which will provide much higher resolution.

An alternative approach is to collide a beam with a sample of
trapped atoms or molecules that are already nearly at rest.
Sawyer {\em et al.} \cite{Sawyer:2008} have recently measured
collision cross sections for He atoms and H$_2$ molecules in
pulsed beams colliding with magnetically trapped OH radicals,
and achieved a resolution of 9 cm$^{-1}$. Experiments to
investigate the collisions of a velocity-controlled beam of
ND$_3$ with trapped ultracold Rb atoms are also under way, and
again should be able to provide much better velocity
resolution.

Scattering resonances in molecular collisions have been studied
for many years \cite{Levine:1968} and can be very important in
chemical reactions \cite{Fernandez-Alonso:2002, Liu:2002}. In
simple systems where only a few partial waves contribute, they
may produce sharp structures in cross sections as a function of
collision energy \cite{Hutson:sbe:1984, Balakrishnan:2000}.
However, in more complicated systems with dense energy level
patterns, the resonant structures may get lost in the
background. It is therefore important to explore whether
well-defined resonant structures are expected for collisions of
molecules that can be velocity-controlled (such as ND$_3$ and
OH) and atoms that can be laser-cooled (such as alkali metal
atoms).

In this paper we study collisions of Rb atoms with ND$_3$ and
NH$_3$ for collision energies between 0 and 100 cm$^{-1}$. We
carry out quantum-mechanical calculations of integral elastic
and state-to-state inelastic cross sections using the coupled
states approximation and observe numerous scattering
resonances. To understand the nature of the scattering
processes, we study the resonance structure of the scattering
cross sections for individual partial waves. The resonances can
be explained in terms of the bound states of the Rb--NH$_3$
complex. We calculate the pattern of bound states near the
lowest dissociation limits of the complex as a function of the
end-over-end angular momentum and identify the bound states
responsible for the resonances.

There is a further reason for being interested in Rb--NH$_3$
collisions. Methods such as Stark deceleration can slow
molecules to velocities of a few metres per second, and the
resulting molecules can then be trapped at temperatures of 10
to 100 mK \cite{Bethlem:trap:2000, Bethlem:tvar:2002}. It has
not yet proved possible to cool such molecules further, towards
the temperatures and phase space densities at which they might
undergo condensation to form quantum gases. There is great
interest in methods that might be used to achieve this, and one
of the most promising is {\em sympathetic cooling}
\cite{Soldan:2004, Lara:PRL:2006}, in which molecules are
cooled by thermal contact with a laser-cooled gas of atoms such
as Rb. However, sympathetic cooling can work only if the
collisions are predominantly elastic rather than inelastic: if
the molecules are in excited states and undergo inelastic
(deexcitation) collisions, the kinetic energy released is
usually enough to eject both collision partners from the trap.
We therefore also carry out calculations of elastic and
inelastic cross sections at the temperatures relevant to
sympathetic cooling.

\section{Theory}
\subsection{Rb--NH$_3$ interaction potential}

In previous work we obtained the potential energy surface for NH$_3$
interacting with Rb from highly correlated electronic structure
calculations \cite{Zuchowski:NH3:2008}. Spin-restricted
coupled-cluster calculations including single, double and
perturbative triple excitations [RCCSD(T)] were carried out on a
grid of points in the intermolecular distance $R$ and intermolecular
angles $\theta$ and $\chi$. The angle $\chi$ corresponds to rotation
of NH$_3$ about its $C_3$ axis. The angles $\theta_i$ were chosen to
be the points for 20-point Gauss-Lobatto quadrature, which include
$\theta=0$ and $180^\circ$ (with $\theta=0$ corresponding to
approach toward the lone pair of NH$_3$). Calculations were carried
out on a grid of $R$ values from 3.5 to 25 $a_0$ and for two
azimuthal angles $\chi=0$ and $60^\circ$.

For collision calculations, we need to evaluate matrix elements
of the interaction potential between angular momentum basis
functions. This is most easily achieved by expanding the
potential in renormalized spherical harmonics $C_{\lambda \mu}
(\theta, \chi)$. After taking account of the $C_{3v}$ symmetry
of NH$_3$, the expansion takes the form
\begin{eqnarray}
V(R,\theta,\chi)  &=&  \sum_{\lambda=0,1,\ldots}
\sum_{\mu=0,3,6,\ldots} \frac{1}{ 1 + \delta_{\mu,0}} v_{\lambda\mu} (R)
\nonumber\\&\times&
\left[ C_{\lambda\mu}(\theta, \chi) + (-1)^\mu C_{\lambda,-\mu} (\theta, \chi) \right] .
\label{pot_exp1}
\end{eqnarray}
The dominant terms in this expansion are those with $\mu=0$ and
3, and the corresponding coefficients may be written
\begin{equation}
v_{\lambda\mu}(R)  =
\frac{ ( \lambda+\frac{1}{2} )}{2 - \delta_{\mu 0} }
\sqrt{ \frac{(\lambda-\mu)! }{(\lambda+\mu)!} }
\int P_{\lambda\mu} (\cos \theta) V_\mu(R,\theta) {\rm d} \cos \theta;
\label{pot_exp2}
\end{equation}
where  $ P_{\lambda\mu} (\cos \theta)$  is an associated
Legendre polynomial and
\begin{eqnarray}
V_0(R,\theta) &=& \textstyle{\frac{1}{2}}[V(R,\theta,0)+
V(R,\theta,60^\circ)] \nonumber \\ V_3(R,\theta) &=&
\textstyle{\frac{1}{2}}[V(R,\theta,0)- V(R,\theta,60^\circ)].
\label{V0V3}
\end{eqnarray}
$V_0$ can be viewed as the interaction potential averaged over
$\chi$, while $V_3$ describes the leading anisotropy of the
potential with respect to rotation about the $C_3$ axis of
NH$_3$. To obtain the expansion coefficients in Eq.\
(\ref{pot_exp2}) at a given value of $R$ for Rb--NH$_3$, the
potential functions $V_\mu(R,\theta_i)$ at each grid point
$\theta_i$ are first evaluated by reproducing kernel Hilbert
space (RKHS) interpolation \cite{Ho:1996, Soldan:2000} and the
angular integrations are then carried out by Gauss-Lobatto
quadrature.

For calculations on Rb--ND$_3$ we need to reexpand the
potential in a coordinate system based on the center of mass of
ND$_3$ instead of NH$_3$. This is done by first generating a
set of points for Rb--ND$_3$ at the same grid of distances and
angles as for Rb--NH$_3$. The center of mass of ND$_3$ is
shifted by a distance $\delta$ towards the D atoms from that of
NH$_3$. Each point is obtained by as
\begin{equation}
V_{\rm Rb-ND_3}(R,\theta,\chi) = V(R',\theta',\chi)
\end{equation}
where
\begin{eqnarray}
R' &=& R(1+t^2+2t\cos\theta)^{1/2}, \nonumber\\
\cos\theta' &=& (\cos\theta+t)R/R',
\end{eqnarray}
and $t=\delta/R$. The potential coefficients for Rb--ND$_3$ are
then obtained by quadrature exactly as for Rb--NH$_3$, but with
the transformed set of points.

\begin{figure}
\includegraphics[width=\linewidth]{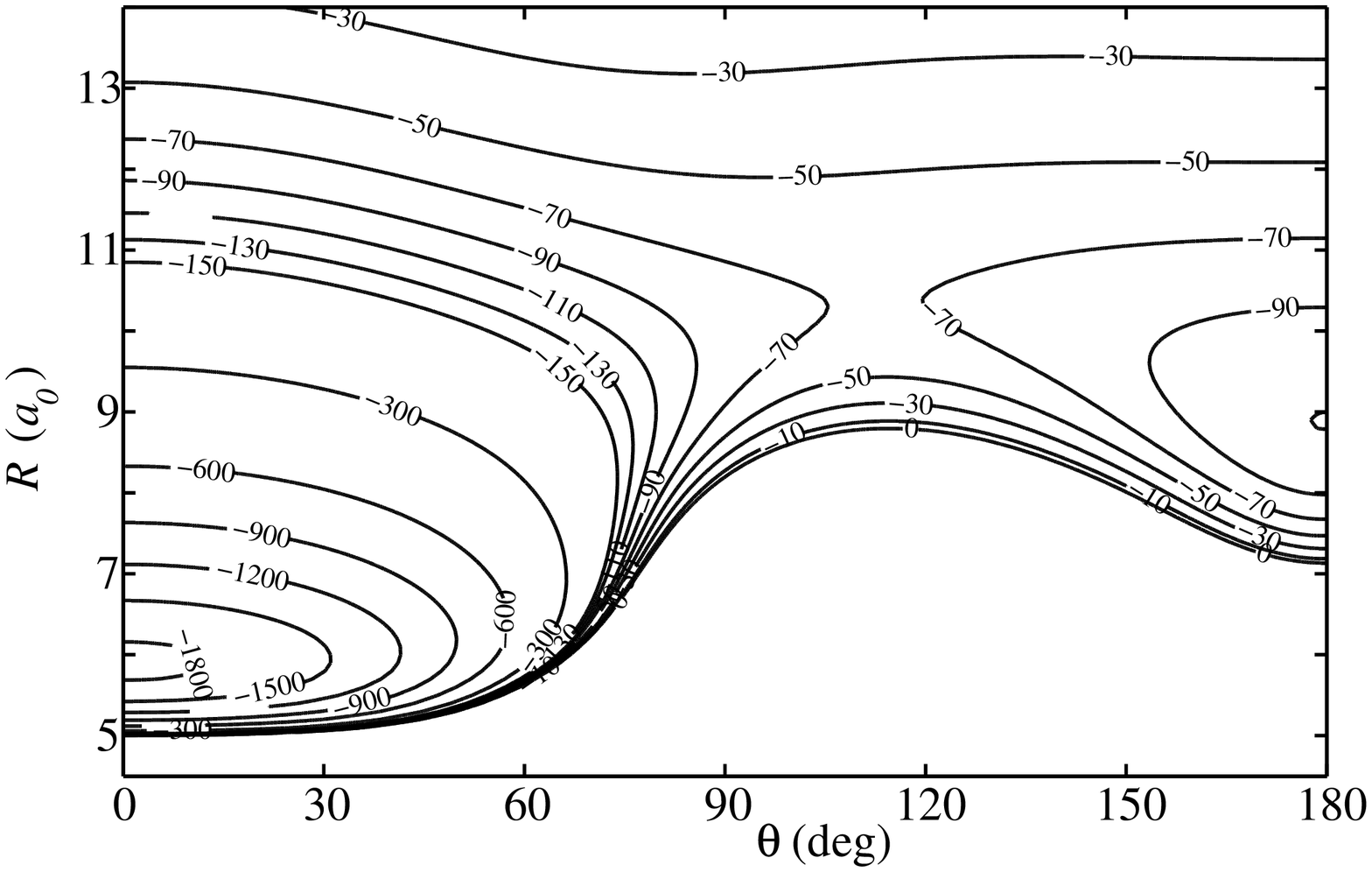}
\includegraphics[width=0.9\linewidth]{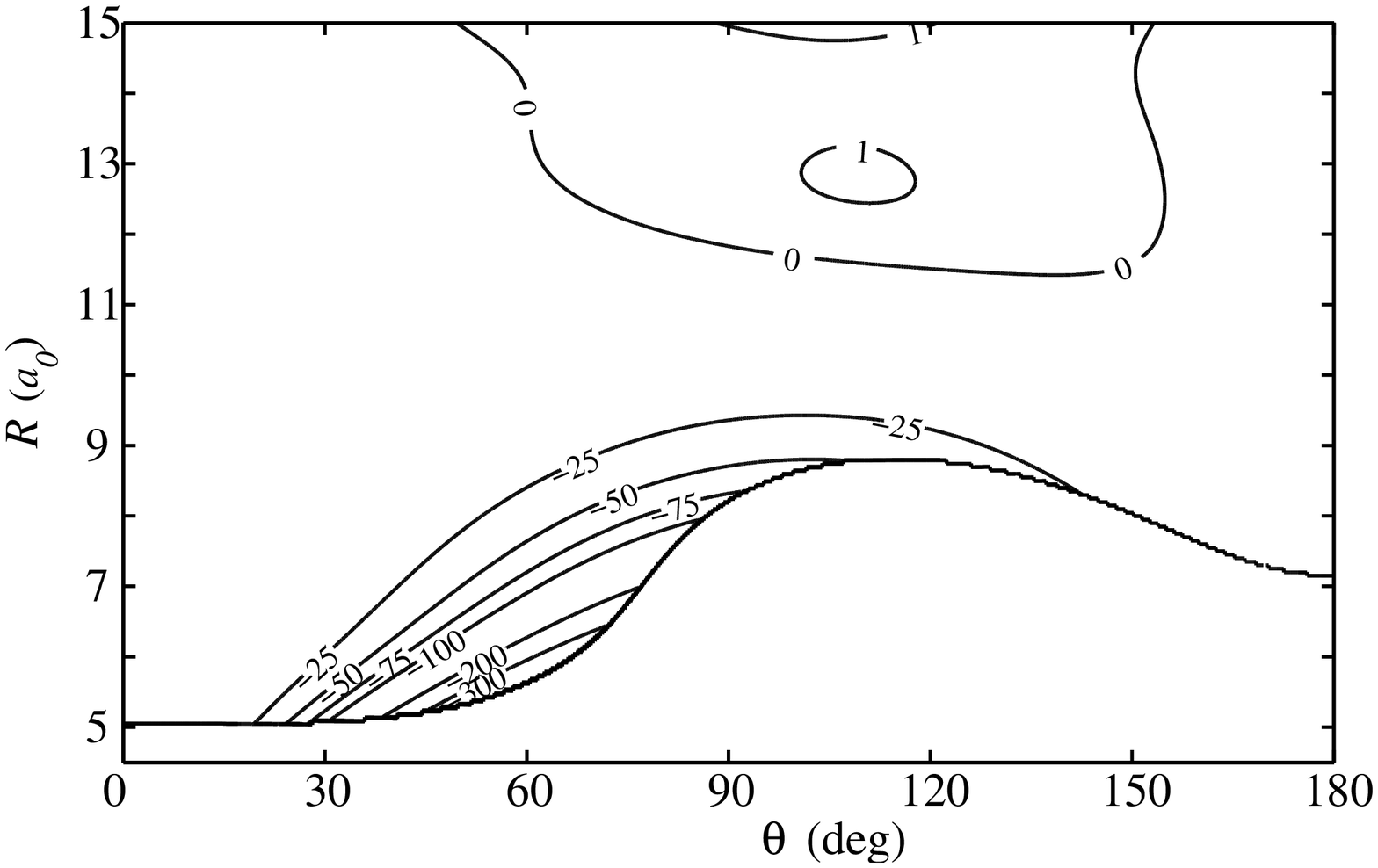}
\caption{The interaction potential of Rb--NH$_3$ from CCSD(T)
calculations: $V_0(R,\theta)$ component (upper panel) and
$V_3(R,\theta)$ component (lower panel). Contours are labeled in
cm$^{-1}$. To aid visualization, $V_3$ is plotted only in the
energetically accessible region defined by $V_0<0$. }
\label{potential}
\end{figure}

The $V_0$ and $V_3$ interaction potentials for Rb--NH$_3$ are
shown in Fig.\ \ref{potential}. The potential is strongly
anisotropic; there is a very deep minimum, nearly 1900
cm$^{-1}$ deep, at the N side of NH$_3$. This is due to the
interaction of the lone pair of NH$_3$ with the singly occupied
$5s$ orbital of the Rb atom. There is also a shallow secondary
minimum, only about 100 cm$^{-1}$ deep, on the H side of
NH$_3$, arising from dispersion interactions. Both minima are
on the $C_3$ axis of NH$_3$. A detailed discussion of the
alkali-atom--NH$_3$ interaction can be found in the Ref.
\cite{Zuchowski:NH3:2008}.

The basis set which we used in our previous work (augmented
triple-zeta for N and H and a basis set of triple-zeta quality for
Rb) gave depths for the minima at linear geometries of 1862 and 110
cm$^{-1}$ (for N-side and H-side configurations, respectively). To
improve this in the present work, we performed additional
calculations for linear geometries near the minima with a
significantly larger basis set. With the basis set extended to
augmented quadruple-zeta for N and H, and with additional $h$
functions on the Rb atom (with exponents 0.45,0.167), and an
additional midbond $g$ function (with exponent 0.4), we obtained a
global minimum well depth of 1881 cm$^{-1}$ with negligible (less
than 0.1\%) change in the equilibrium distance. All the potential
points were then scaled by the ratio of the well depths calculated
with quadruple-zeta and triple-zeta basis sets. The depth of the
secondary minimum on the surface obtained by such scaling is 111
cm$^{-1}$, which agrees well with the value obtained from {\em ab
initio} calculations with the quadruple-zeta basis set. In the
scattering calculations described below we have used the scaled
triple-zeta potential.

For very large Rb--NH$_3$ separations the expanded potential is
dominated by the $v_{0,0}$ and $v_{2,0}$ terms, which vanish as
$R^{-6}$. Since the inherent error of the supermolecular
approach is quite significant at large $R$, we represent the
long-range behavior of $v_{0,0}$ and $v_{2,0}$ using $C_{6,0}$
and $C_{6,2}$ coefficients obtained from perturbation theory
\cite{Zuchowski:NH3:2008}. At short range we switch over to the
full expansion using a switching function
$\{1+\mbox{exp}[a(R-b)] \}^{-1}$ with $a=0.5\ a_0^{-1}$ and
$b=26\ a_0 $.

\subsection{Scattering calculations}

The Hamiltonian that describes collisions between an atom and a
molecule may be written
\begin{equation}
\frac{\hbar^2}{2\mu} \left[-\frac{d^2}{dR^2} + \frac{\hat L^2}{R^2}\right]
+ H_{\rm at} + H_{\rm mol}
+ V_{\rm inter},
\end{equation}
where $\mu$ is the reduced mass for the colliding pair, $\hat L^2$
is the operator for rotation of the collision partners about one
another, and $H_{\rm at}$ and $H_{\rm mol}$ are the Hamiltonians for
the isolated atom and molecule respectively. In the special case
where either the atom or the molecule is in a closed-shell singlet
state, the main part of the interaction potential $V_{\rm inter}$ is
a single potential energy surface $V(R,\theta,\chi)$ that couples
the molecular rotations strongly to the intermolecular distance. The
molecular rotations are in turn coupled to the nuclear spins in the
molecule by hyperfine terms. However, for an atom in an S state the
only terms that couple the molecular degrees of freedom to the {\em
atomic} electron and nuclear spins are (i) very weak magnetic
dipole-dipole interactions between atomic and molecular spins and
(ii) any dependence of the atomic hyperfine coupling on the
intermolecular distance $R$. If these very small terms are
neglected, the collision problem may be treated as the scattering of
the molecule from an {\em unstructured} atom.

The Hamiltonian of a rigid symmetric-top molecule may be
written
\begin{equation}
H_{\rm mol} = B \hat{j}^2 + (C-B) \hat{j}_{z}^2
\label{symtophamiltonian}
\end{equation}
where $\hat{j}^2 $ is the operator representing the angular
momentum of the molecule and $\hat{j}_{z}$ denotes its
projection onto the molecule-fixed symmetry axis $z$, while $C$
and $B$ are the rotational constants for rotation about the
symmetry axis and an axis perpendicular to it. The
eigenfunctions of this Hamiltonian are labeled by three quantum
numbers: $j$, the total angular momentum of the rotor; $m$, the
projection of $j$ on the laboratory-frame $Z$-axis and $k$, the
projection of $j$ on the body-fixed molecular symmetry axis
$z$,
\begin{eqnarray}
\hat{j}^2 | j km \rangle &=& j(j+1) |jkm \rangle \nonumber \\
\hat{j}_Z | j km \rangle &=& m |jkm \rangle \nonumber \\
\hat{j}_z | j km \rangle &=& k |jkm \rangle.
\label{eigeq}
\end{eqnarray}

The N-H stretching vibrations are at sufficiently high
frequency that they can be neglected in the present work, but
NH$_3$ also has a bending (umbrella) vibration with two minima
that are equivalent in the free molecule. Tunnelling between
these two minima produces a low-frequency splitting that
corresponds to inversion of the molecule through a planar
geometry. To describe this we introduce an additional degree of
freedom, the inversion coordinate $h$. The vibration-inversion
functions $|+ \rangle$ and $|-\rangle$ (corresponding to the
lower and upper component of the inversion doublet) may be
written
\begin{equation}
| \pm \rangle \sim
 \left[ f(h_{\rm eq} - h )  \pm f(h_{\rm eq} + h ) \right],
\end{equation}
where $f(x)$ is a function that is sharply peaked around $x=0$,
so that $f(h_{\rm eq} \pm h )$ is a bending function that is
sharply peaked around one of the equilibrium values $\mp h_{\rm
eq}$ (corresponding to one of the individual ``umbrella''
states of NH$_3$).

NH$_3$ occurs in two different nuclear spin configurations,
corresponding to $A_1$ and $E$ symmetries. In the first case
(referred to as ortho-NH$_3$) the molecule can occupy
rotational levels with $k=0, 3, 6,\ldots$, while in the latter
(referred to as para-NH$_3$) only $k=1,2,4,5\ldots$ are
allowed. For ortho-NH$_3$ in $k=0$ states, only one component
of each inversion doublet is allowed by symmetry. ND$_3$ is
slightly different and occurs in three nuclear spin symmetries,
$A_1$, $A_2$ and $E$. Molecules in nuclear spin states of $A_1$
and $A_2$ symmetries can occupy rotational levels with $k=0$,
3, 6, etc., referred to as ortho-ND$_3$, while those in the $E$
state can occupy levels with $k=1$, 2, 4, 5, etc., referred to
as para-ND$_3$. Ortho- and para-species do not readily
interconvert in collisions. By contrast with NH$_3$, however,
ND$_3$ molecules in $k=0$ states can exist in both upper and
lower components of the inversion doublet.

The rotational functions adapted for permutation-inversion
 symmetry  may be written \cite{Green:1980}
\begin{equation}
\left[ | j k m \rangle \mp (-1)^j |j -\!\!k m \rangle \right] | \pm \rangle
\end{equation}
for para-NH$_3$ and
\begin{equation}
[ | j k m \rangle \pm (-1)^j |j -\!\!k m \rangle ] | \pm \rangle
\end{equation}
for para-ND$_3$. If $f(h_{\rm eq} - h )$ and $f(h_{\rm eq} + h
)$ do not overlap significantly, evaluating of the matrix
elements of the potential between the symmetry-adapted basis
functions gives expressions that are isomorphic to those for
the case of rigid symmetric top with parity-adapted monomer
basis functions. It is therefore sufficient to carry out the
scattering calculation exactly as for a rigid symmetric top,
but with the monomer energies for the even and odd combinations
of $| j k m \rangle$ and $|j -\!\!k m \rangle$ shifted up and
down from the rigid-rotor values by half the inversion
splitting.

The approach of restricting the basis set of inversion
functions to just the lowest pair of states was introduced by
Green \cite{Green:1976, Green:1980} and Davis and Boggs
\cite{Davis:1978}, but was later verified to be a good
approximation by Van der Avoird and coworkers
\cite{vanderSanden:1992, vanBladel:1991}, who carried out
calculations on Ar--NH$_3$ collisions to compare with theory
that took the $h$ coordinate into account explicitly.

The rotational constants and inversion parameters used for
NH$_3$ and ND$_3$ in the present paper are given in Table
\ref{molpar}. Since inelastic transitions between inversion
states are of key interest for the present work, we allow the
inversion splitting to be a function of rotational state
\cite{Tow55},
\begin{equation}
\nu = \nu_0 - \nu_a [j(j+1) -k^2]  - \nu_b k^2
\end{equation}
The energy levels for $j=1$ to 3 for both ortho- and
para-NH$_3$ and ND$_3$ are given in Table \ref{enlevels}. In
each case the zero of energy is the energy of the lowest
allowed tunneling component for $j=0$, $k=0$.

In the following sections we identify tunneling
components by $u$ (upper) or $l$ (lower) instead of symmetry
labels, because the $u/l$ designation indicates the energetics
more directly.

\begin{table}
\caption{Rotational constants \cite{Poynter:1983, Urban:1984}
and inversion splitting parameters (with centrifugal
distortions) \cite{Tow55} used for NH$_3$ and ND$_3$. Units are
cm$^{-1}$.} \label{molpar}
   \centering
   \begin{ruledtabular}
   \begin{tabular}{lll}
    parameter   &  NH$_3$                           &   ND$_3$                \\ \hline
    $B$         &  9.9441                           &  5.1428                 \\
    $C$         &  6.2294                           &  3.1142                 \\
    $\nu_0$     &  0.7934                           &  5.337$\times 10^{-2}$    \\
    $\nu_a$     &   5.05 $\times 10^{-3}$           &   $2.39\times10^{-4}$   \\
    $\nu_b$     &   1.998$\times 10^{-3}$           &   $9.61 \times 10^{-5}$ \\
   \end{tabular}
\end{ruledtabular}
\end{table}

\begin{table}
\caption{The energy levels of NH$_3$ and ND$_3$ molecules (up
to $j=3$) used in the present calculations. Units are
cm$^{-1}$. The labels o and p refer to ortho- and para- spin
isomers. For NH$_3$ in $k=0$ states, only the
$|(-1)^{j+1}\rangle$ inversion state is allowed for each $j$.}
\label{enlevels}
   \centering
   \begin{ruledtabular}
   \begin{tabular}{ccrcrc}
   state       & spin isomers    &  NH$_3$           & $|\pm\rangle$   &  ND$_3$    & $|\pm\rangle$  \\ \hline
    $00$       & o     &        0.0000              & $-$   &       0.0000       &  $+$    \\
    $00$       & o     &                            &       &       0.0534       &  $-$    \\
    $11$       & p     &        15.3871             & $+$   &       8.2570       &  $+$    \\
    $11$       & p     &        16.1735             & $-$   &       8.3100       &  $-$    \\
    $10$       & o     &        19.1049             & $+$   &     10.2851      &  $+$    \\
    $10$       & o     &                            &       &     10.3385      &  $-$    \\
    $22$       & p     &        44.0305             & $+$   &       22.7424      &  $+$    \\
    $22$       & p     &        44.8058             & $-$   &       22.7949      &  $-$    \\
    $21$       & p     &        55.1837             & $+$   &       28.8282      &  $+$    \\
    $21$       & p     &        55.9499             & $-$   &       28.8802      &  $-$    \\
    $20$       & o     &        59.6646             & $-$   &       30.8568      &  $+$    \\
    $20$       & o     &                            &       &       30.9102      &  $-$    \\
    $33$       & o     &        85.1365             & $+$   &   43.4562          & $+$     \\
    $33$       & o     &        85.8969             & $-$   &   43.5080          & $-$     \\
    $32$       & p     &       114.8786             & $+$   &   59.6850          & $+$     \\
    $32$       & p     &       115.6145             & $-$   &   59.7356          & $-$     \\
    $31$       & p     &       103.7254             & $+$   &   53.5992          & $+$     \\
    $31$       & p     &       104.4704             & $-$   &   53.6503          & $-$     \\
    $30$       & o     &       118.5964             & $+$   &   61.7107          & $+$    \\
    $30$       & o     &                            &       &   61.7641          & $-$   \\
   \end{tabular}
\end{ruledtabular}
\end{table}

The Rb--NH$_3$ potential is very strongly anisotropic. Because
of this, scattering calculations require large basis sets of
rotational functions for convergence. In addition, we typically
need to carry out calculations for many different total angular
momenta (partial waves) to obtain all the contributions to
integral cross sections. We typically need to include 100 to
200 partial waves at collision energies around 100 cm$^{-1}$.

The most accurate approach for quantum inelastic scattering is
to use close-coupling calculations, which expand the total
wavefunction for the collision system in a space-fixed basis
set using a total angular momentum representation. In the
present case the basis functions are labeled by quantum numbers
$j,k,\pm$, which describe the monomer, and $L$, which describes
the angular momentum for rotation of the collision partners
about one another. For each total angular momentum $J$ and
total parity, full close-coupling calculations include all
possible values of $L$ allowed by angular momentum coupling for
each monomer level $(j,k,\pm)$. However, such calculations give
enormous basis sets for large $J$, and are prohibitively
expensive except at the very lowest energies (where only a few
$J$ values contribute to cross sections).

A more affordable approach that is adequate at higher energies
is provided by the coupled-states (CS) approximation
\cite{McG74,Pac74}, which was introduced for atom-symmetric top
systems by Green \cite{Green:1976}. In the CS approach the
scattering equations are written in a body-fixed frame and for
each partial wave the basis functions are labeled by the
monomer quantum numbers $(j,k,\pm)$ and the projection $K$
(helicity) of $j$ onto the body-fixed intermolecular axis $R$.
If the centrifugal operator is approximated by $L(L+1)/(2\mu
r^2)$, with $L=J$ for all channels, then the coupled equations
factorize into independent sets labeled by $L$ and $K$. The
size of the resulting basis sets is independent of $L$, so that
the total cost of a scattering calculation scales linearly with
the number of partial waves included. The CS approximation is
generally expected to be accurate when the kinetic energy and
the potential anisotropy are both large compared to the
rotational constant of the collision complex.

\subsection{Details of scattering calculations}

We carry out coupled-channel scattering calculations using the
MOLSCAT package \cite{molscat:v14} with the hybrid
log-derivative/Airy propagator of Manolopoulos and Alexander
\cite{Alexander:1987}. The inner starting point for the
integration was chosen to be $R=4.2 \ a_0$, which is deep in
the inner classically forbidden region. Since the potential for
$\theta \approx 0^\circ$ is very deep, the wavefunction
oscillates very rapidly at short range. Thus in the inner
region ($R<13\ a_0$) we use the fixed-interval log-derivative
propagator \cite{Manolopoulos:1986} with a small step size of
0.015 $a_0$. The log-derivative matrix is then propagated from
$13 \ a_0$ to 400 $a_0$ with the variable-step Airy propagator,
which takes very long steps at long range and consumes only a
small part of the total computer time. The Airy propagator was
used with convergence criterion TOLHI = $10^{-4}$
\cite{Alexander:1984}. The resulting inelastic and elastic
cross sections are converged with respect to all integration
parameters to better than 1\%.

As mentioned above, the strong anisotropy of the potential
energy surface causes slow convergence with respect to the
basis set of monomer rotational functions, especially for
ND$_3$, which has rotational constants about a factor of 2
smaller than NH$_3$. In Table \ref{convergence} we show the
convergence of selected integral cross sections for Rb--ND$_3$,
together with relative computer times. The rotational basis set
with $j_{\rm max}=21$ and $k_{\rm max}=8$, containing 210
monomer energy levels for para-ND$_3$, gives a good compromise
between accuracy and computational cost: typical calculations
for a single energy and $K$ value take $\sim 30$ sec on a
2.0-GHz single-core Opteron processor, which means that
calculations of cross sections converged with respect to the
partial-wave expansion take approximately 2 hours per energy.
For Rb--NH$_3$, a slightly smaller basis set with $j_{\rm
max}=14$ and $k_{\rm max}=7$ gave cross sections converged to
$\sim 1\%$.

All calculations are for $^{87}$Rb.

\begin{table}
\caption{The convergence of typical cross sections (in \AA$^2$)
for Rb--ND$_3$ with respect to the maximum quantum numbers
$(j_{\rm max},k_{\rm max})$ in the rotational basis set. The
cross sections were calculated with the coupled-states method
at an energy of 25 cm$^{-1}$ with respect to $j=0,k=0$ level.
The computer time taken for a single partial wave, relative to
the time for the (21,8) basis set, is also given.}
\label{convergence}
\begin{ruledtabular}
\begin{tabular}{lrrrr}
 Basis &  $\sigma_{11u  \to 11 u}$  &  $\sigma_{11u  \to 11l } $
 & $\sigma_{ 22u  \to 11u }$ & time    \\ \hline
  (15,5)  &   1309  &  62.22    &   5.14   & 0.2 \\
  (18,5)  &   1292  &  59.42    &   3.89   & 0.3 \\
  (21,5)  &   1283  &  62.67    &   5.59   & 0.5 \\
 \hline
  (15,8)  &   1296  &  57.39    &   4.94   & 0.3 \\
  (18,8)  &   1291  &  63.11    &   7.45   & 0.6 \\
  (21,8)  &   1293  &  59.70    &   3.90   & 1.0 \\
 \hline
  (15,11) &   1302  &  61.83    &   4.92   & 0.5 \\
  (18,11) &   1291  &  62.09    &   5.77   & 0.9 \\
  (21,11) &   1292  &  59.42    &   3.89   & 1.7 \\
\end{tabular}
\end{ruledtabular}
\end{table}

\section{Results}
\subsection{Scattering cross sections}

\begin{figure}[tbph]
\includegraphics[width=0.9\linewidth]{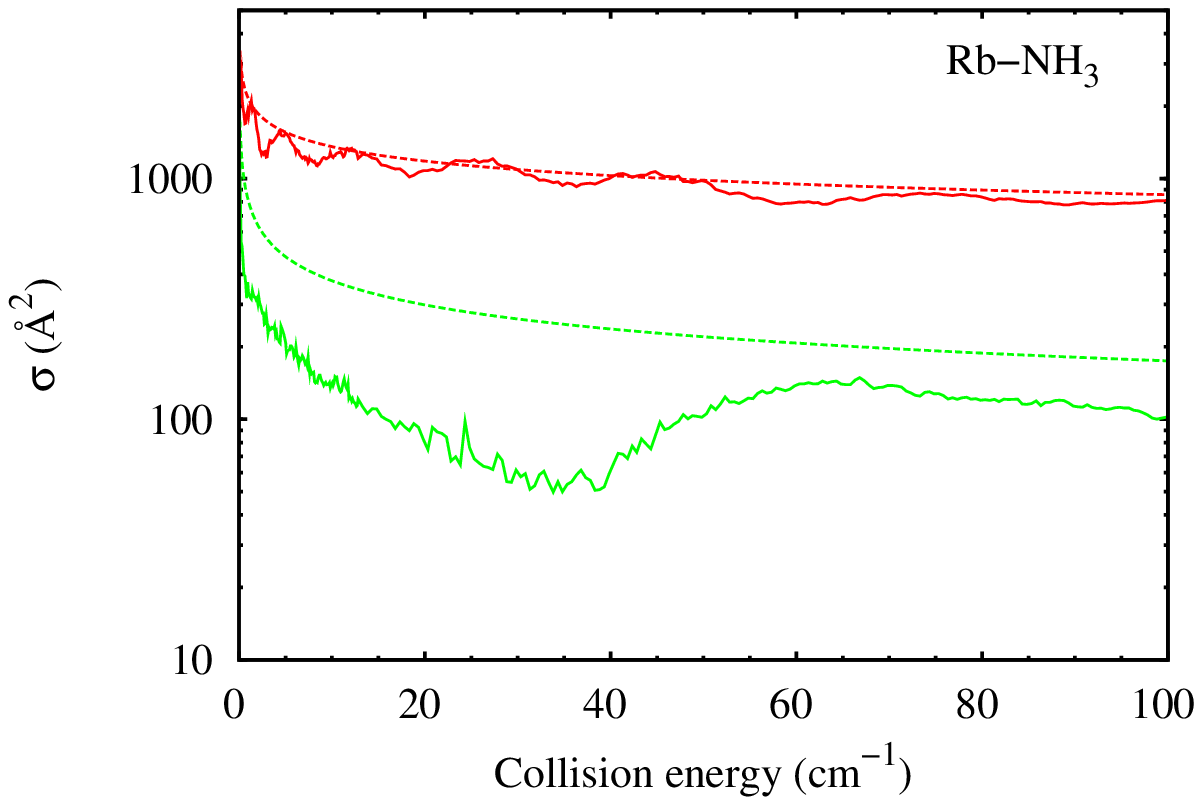}
\includegraphics[width=0.9\linewidth]{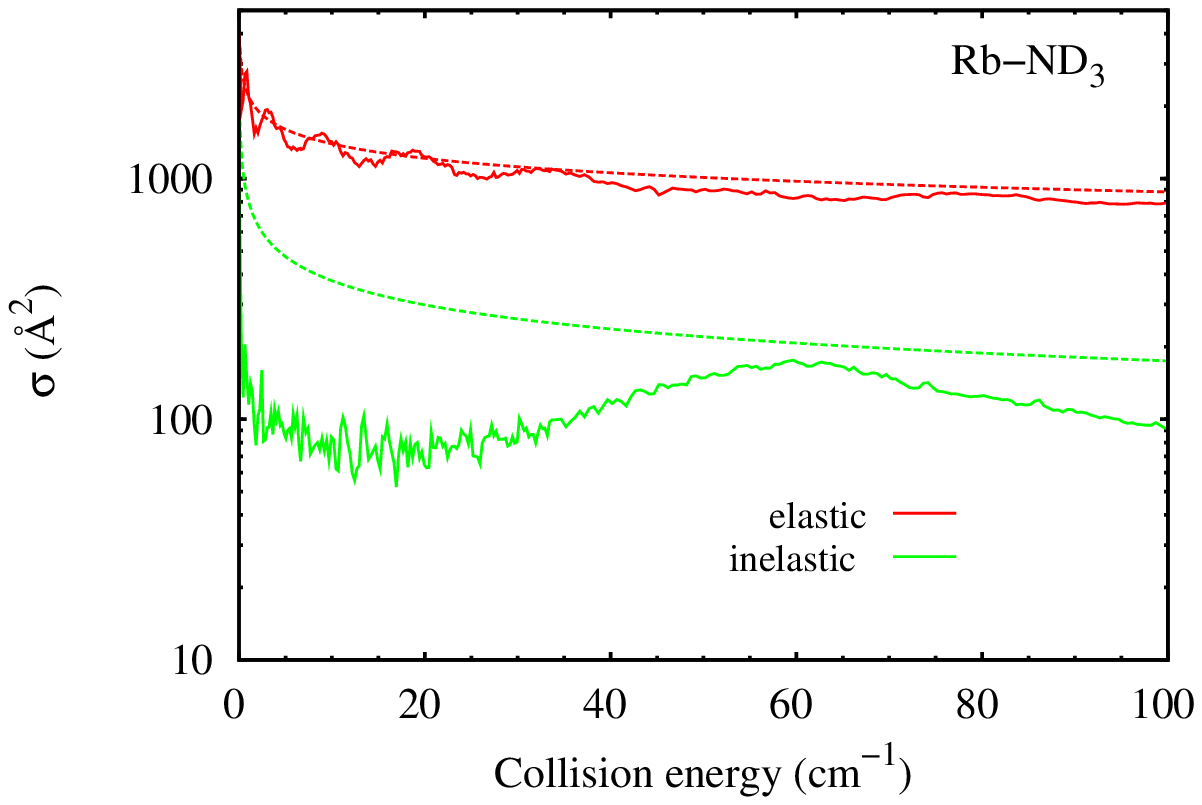}
\caption{Elastic and total inelastic cross sections for
Rb--NH$_3$ (upper panel) and Rb--ND$_3$ scattering (lower
panel) from the upper component of the inversion doublet for
$j=1$, $k=1$ state. The smooth dashed lines show the result of
the semiclassical background formula for elastic cross sections
and of Langevin capture theory for the total inelastic cross
sections. } \label{totalcs}
\end{figure}

The ND$_3$ and NH$_3$ molecules can be slowed by Stark
deceleration in their low-field-seeking $j=1$, $k=1$ states,
which correlate at low field with the upper level of the
inversion doublet. Since ND$_3$ has a much smaller tunneling
splitting than NH$_3$, its Stark effect is more nearly linear
and it is easier to decelerate. We have calculated the energy
dependence of the state-to-state integral cross sections for
molecules initially in the upper inversion state of ND$_3$ and
NH$_3$ for collision energies between 0 and 100 cm$^{-1}$. The
results are shown in Fig.\ \ref{totalcs}.

For the whole energy range considered, the elastic cross
sections $\sigma_{11u \to 11u}$ are large compared to the total
inelastic cross sections for both ND$_3$ and NH$_3$. Fig.\
\ref{totalcs} includes the elastic cross section obtained from
the semiclassical background formula \cite{Child:1991:p212} for
a pure $R^{-6}$ potential (dashed red line). For both
Rb--NH$_3$ and Rb--ND$_3$ the elastic cross sections follow the
background formula fairly closely, with slow glory oscillations
superimposed on the background. The total cross sections
(elastic + inelastic) are actually in even better agreement
with the semiclassical model than the elastic cross sections.
There is also structure due to scattering resonances at low
collision energies, but it is not very strong with respect to
the background for the elastic cross sections. At higher
collision energies the resonances are lost in the background.

\begin{figure}[tbph]
\includegraphics[width=0.9\linewidth]{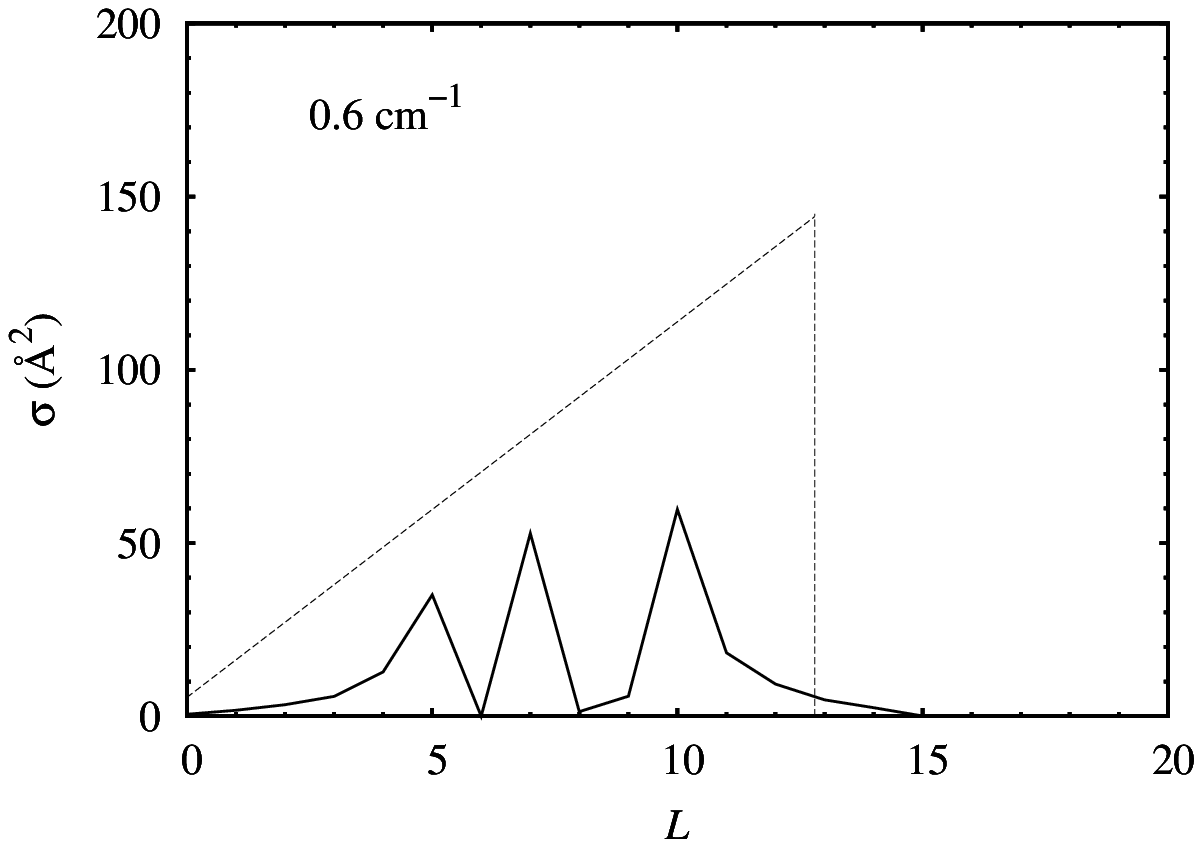}
\includegraphics[width=0.9\linewidth]{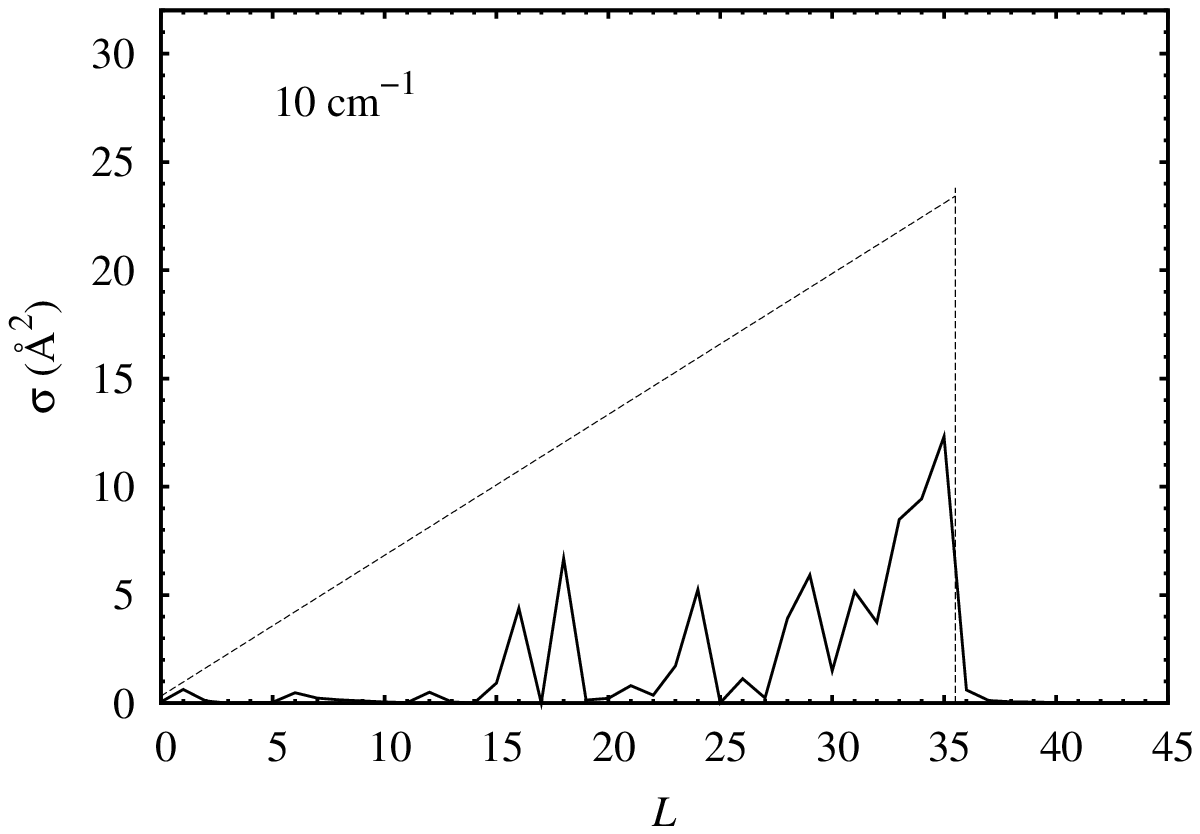}
\caption{Partial-wave contributions to inelastic $11u\to11l$
cross sections for Rb--ND$_3$ at collision energies of 0.6
cm$^{-1}$ (upper panel) and 10 cm$^{-1}$ (lower panel). The
straight dashed lines show the result corresponding to Langevin
capture theory, with a cutoff determined by the heights of the
centrifugal barriers calculated for the full potential.}
\label{opacity}
\end{figure}

The overall magnitude of the inelastic cross sections is at
first sight less than expected. In a strongly coupled system
where every collision that crosses the centrifugal barrier
leads to inelasticity, the total inelastic cross section is
given by the Langevin capture formula \cite{Levine:1987},
\begin{equation}
k_{\rm capture}^{\rm inel} = 3\pi \left(\frac{C_6}{4E}\right)^{1/3}.
\end{equation}
The Langevin result is shown as a dashed green line in Fig.\
\ref{totalcs} and it may be seen that the actual $11u\to 11l$
inelastic cross section is substantially below it. This is
surprising in view of the very large anisotropy which directly
couples the $11l$ and $11u$ states. Further insight is given by
the partial-wave contributions to the $11u\to 11l$ cross
sections, which are shown for collision energies of 0.6
cm$^{-1}$ and 10 cm$^{-1}$ in Fig.\ \ref{opacity}. In this case
the capture cross section corresponds to an elastic S-matrix
element of zero for every $L$ below a cutoff due to the
centrifugal potential. The partial inelastic cross section is
far less than the capture value across most of the range of
$L$. This contrasts with the behavior observed in other
strongly coupled systems, such as high-energy HF--HF collisions
\cite{Alexander:1980}, rotationally inelastic collisions in
Ar--N$_2$ at 300~K \cite{Connor:1990}, and barrierless
reactions in alkali metal atom + diatom collisions at energies
above 1~mK \cite{Cvitas:bosefermi:2005, Quemener:2005,
Cvitas:li3:2007, Hutson:IRPC:2007}, where the partial cross
sections {\em are} close to the capture value for all $L$.

\begin{figure}[tbph]
\includegraphics[width=0.95\linewidth]{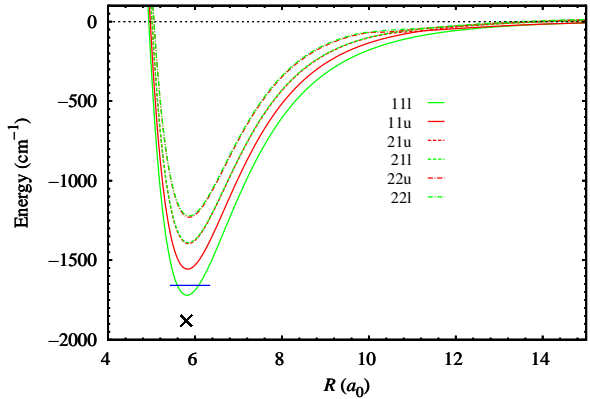}
\caption{Adiabatic bender curves for Rb interacting with
para-ND$_3$. The line types indicate the level that each curve
correlates with at long range and not the character of the adiabatic
state at short range. The cross indicates the position of the
absolute minimum of the potential surface and the horizonal line
indicates the energy of the lowest bound state, $-1658.252$
cm$^{-1}$.} \label{adiabats}
\end{figure}

A possible explanation for the low inelasticity for Rb--ND$_3$ is
that, despite the large anisotropy, the collisions are approximately
adiabatic in the rotation-inversion coordinates. Fig.\
\ref{adiabats} shows ``adiabatic bender" curves for Rb--ND$_3$,
which are eigenvalues of the rotation-inversion Hamiltonian for
para-ND$_3$ at fixed $R$. It may be seen that the curves correlating
with the $11u$ and $11l$ states stay far apart for all values of
$R$, with no avoided crossings between them where strong
nonadiabatic transitions would be expected.

\begin{figure}[tbph]
\includegraphics[width=0.95\linewidth]{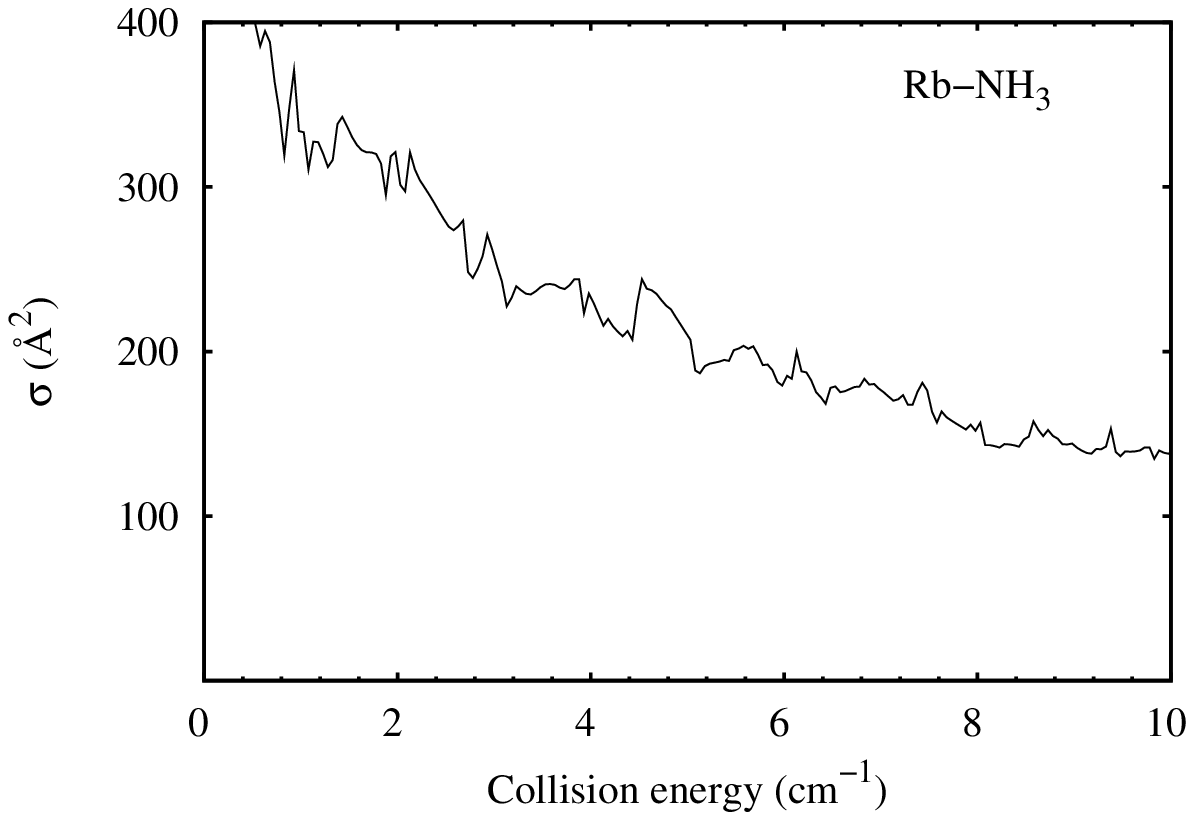}
\includegraphics[width=0.95\linewidth]{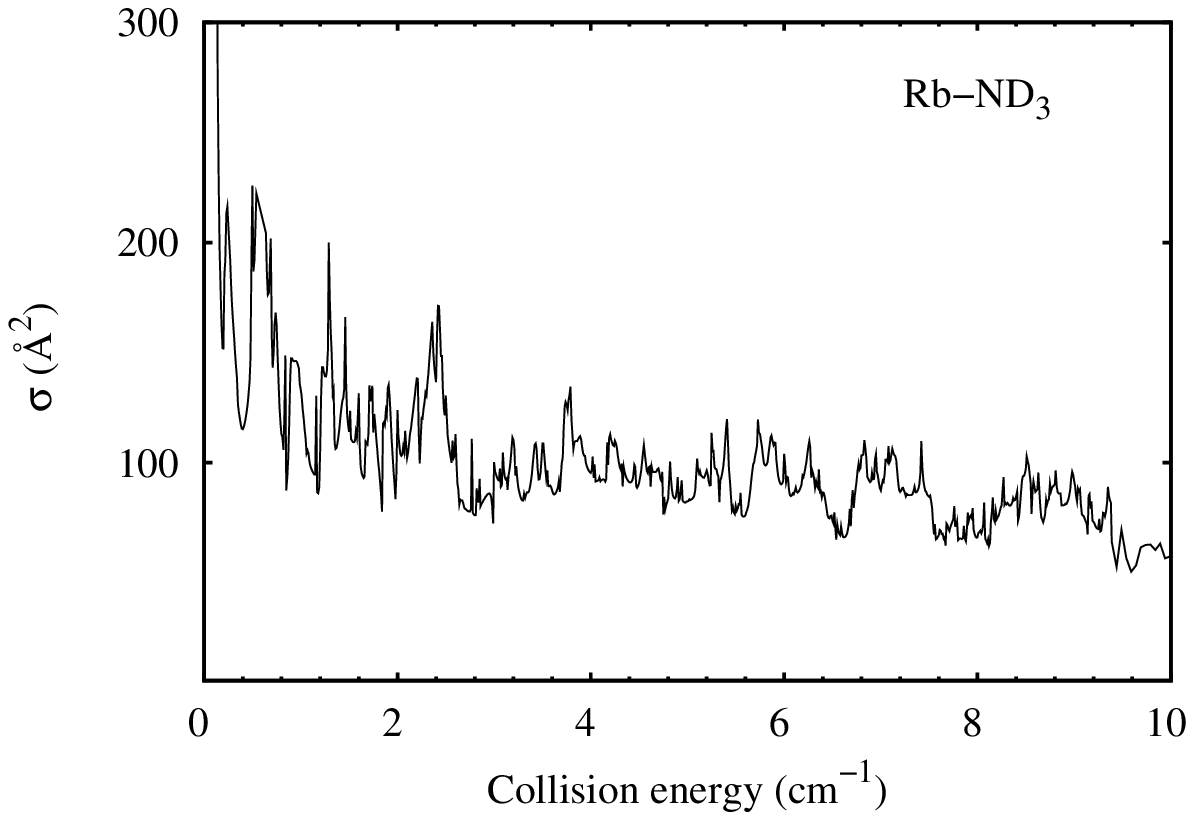}
\caption{Resonances in the inelastic $11u \to 11l$ cross
sections for Rb--NH$_3$ and Rb--ND$_3$ at low collision
energies.} \label{totalcszoom}
\end{figure}

The resonances are much stronger in the inelastic cross
sections than the elastic cross sections. They are particularly
strong for collision energies up to about 20 cm$^{-1}$, as can
be seen in the expanded view of the energy dependence of
$\sigma_{11u \to 11l}$ in Fig.\ \ref{totalcszoom}. For
Rb--ND$_3$ at low energies the resonances can enhance inelastic
cross sections by up to a factor of 2. As the collision energy
increases the resonances again become weaker and wash out. We
see significantly more resonances for Rb--ND$_3$ than for
Rb--NH$_3$. A detailed description of the origin of the
resonances for low kinetic energies will be given in section
III.B.

\begin{figure}[tbph]
\includegraphics[width=0.95\linewidth]{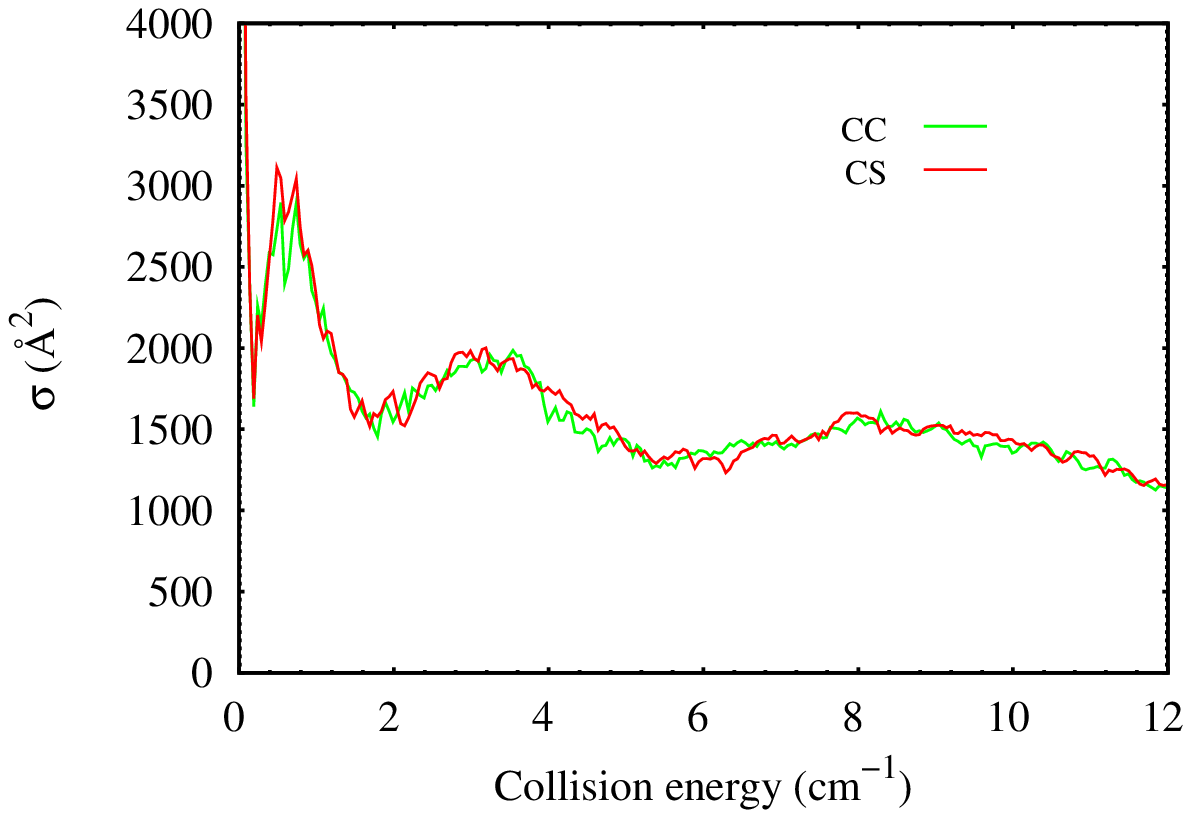}
\includegraphics[width=0.95\linewidth]{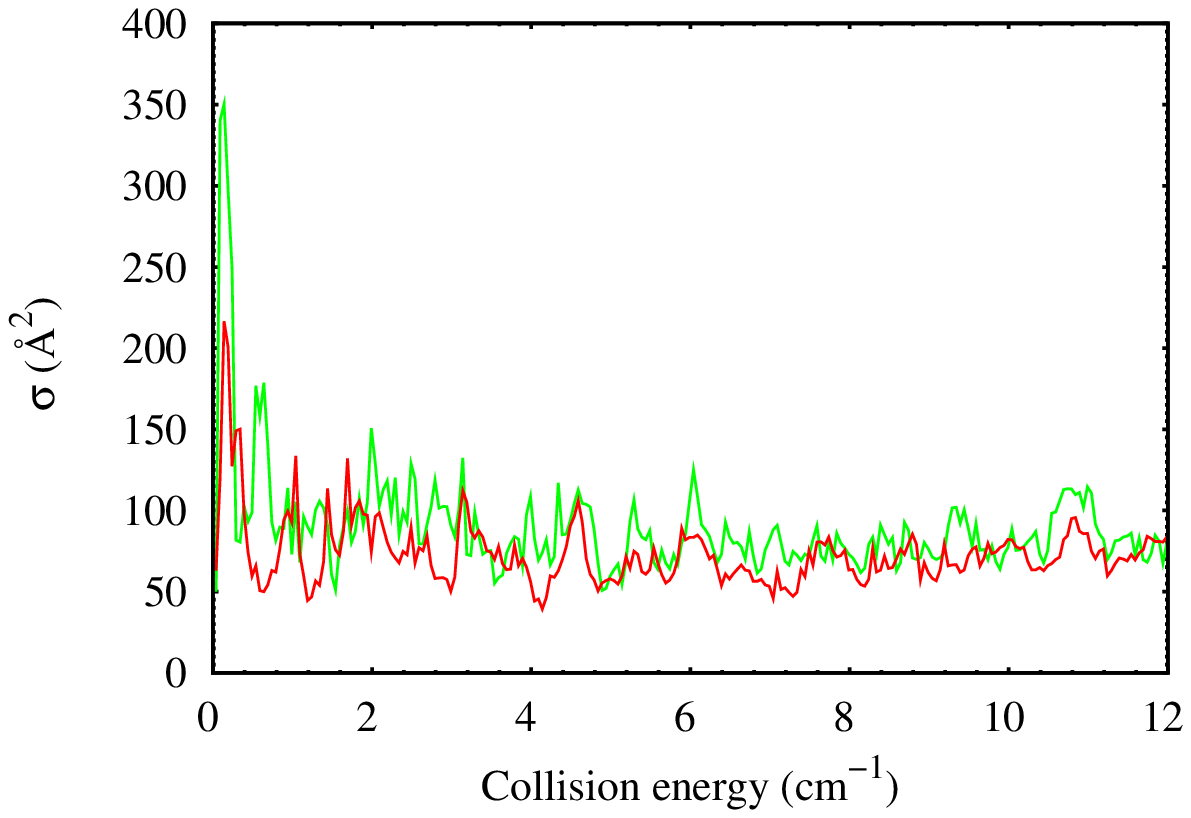}
\caption{Comparison of elastic $11u \to 11u$ (upper panel) and
inelastic $11u \to 11l$ (lower panel) cross sections for
Rb--ND$_3$ calculated with CS and CC methods for a restricted
basis set with $j_{\rm max}=6$, $k_{\rm max}=5$.}
\label{CCvsCS}
\end{figure}

The coupled states approximation is not expected to be accurate
at very low energies, because it approximates the centrifugal
terms in the potential and neglects couplings between different
helicities. Unfortunately, full close-coupling calculations for
higher energies are prohibitively expensive for partial waves
corresponding to large values of $J$. For example, the basis
set used above ($j_{\rm max}=21$ and $k_{\rm max}=8$), which
gives 210 channels for para-ND$_3$ in the CS approximation,
produces 2745 channels for each parity in CC calculations for
$J\ge21$. However, it is possible to make the comparison for a
restricted basis set, and the results are shown in Fig.\
\ref{CCvsCS} for a basis set with $j_{\rm max}=6$, $k_{\rm
max}=5$. The terms that are neglected in the CS approximation
cause small shifts in resonance positions and intensities, so
that it is not possible to make a valid comparison at any
individual collision energy. However, the general extent of the
resonance structure is similar in the two calculations and the
inelastic cross sections typically agree within 5\% above 30
cm$^{-1}$. We can therefore have confidence in the general
features of the CS calculations for the full basis set.

\begin{figure}[tbph]
\includegraphics[width=0.95\linewidth]{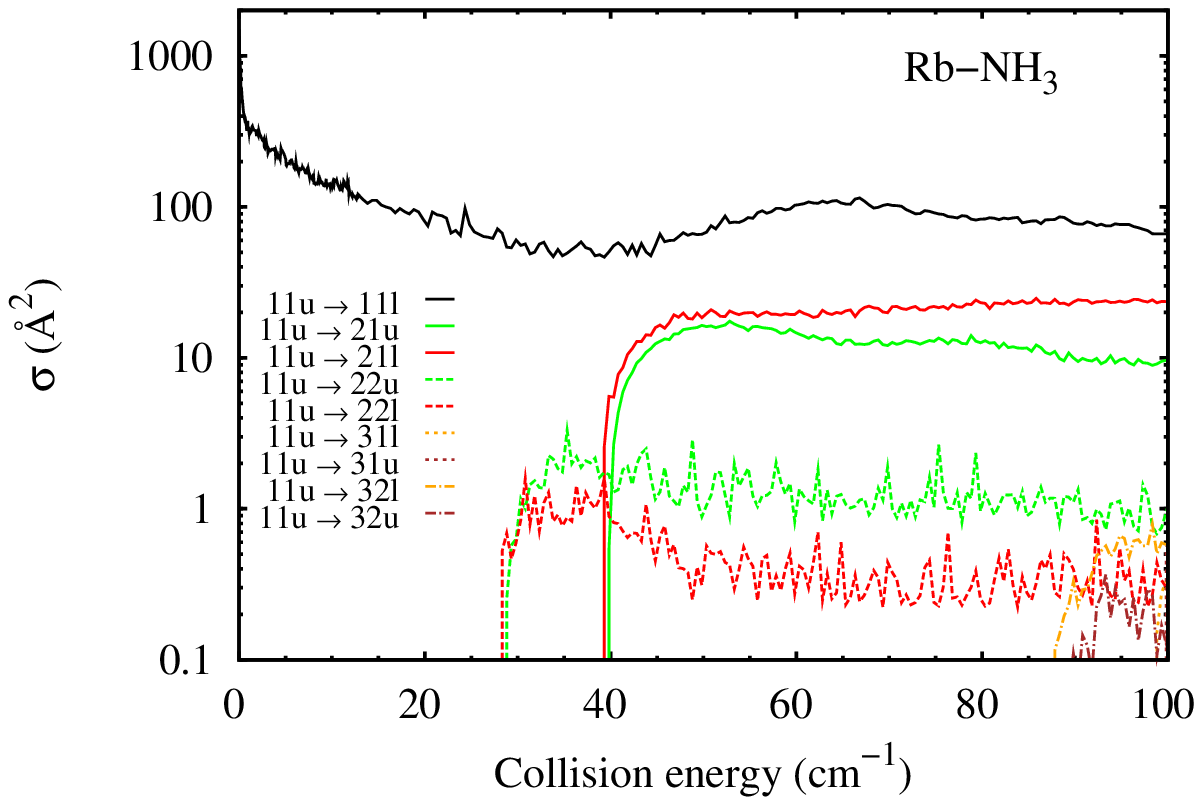}
\includegraphics[width=0.95\linewidth]{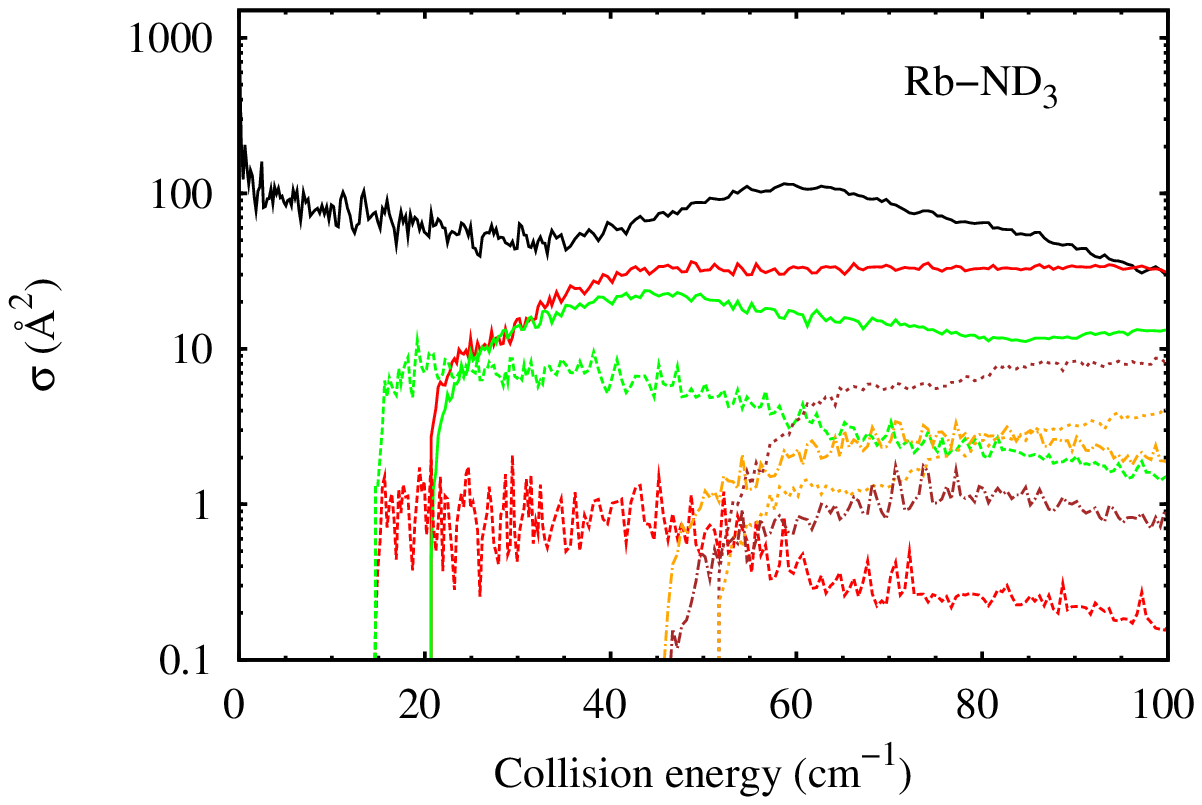}
\caption{State-to-state inelastic cross sections from the $11u$
state of Rb--NH$_3$ and Rb--ND$_3$ as a function of collision
energy, from coupled-states calculations.} \label{inelcs}
\end{figure}

The state-to-state cross sections from the $11u$ state to other
$j=1\ldots3$ states are shown as a function of collision energy
in Fig.\ \ref{inelcs}. As the transition to each additional
rotational state becomes energetically allowed, the
corresponding excitation cross section rises sharply from zero.
This behavior could be seen readily in experiments that use
state-selective detection for the scattered ND$_3$. However,
since the cross sections for rotational excitation are very
small compared to those for $11u \to 11l$ relaxation at these
energies, we do not see any significant changes in the total
inelastic cross section as new channels open up.

\begin{figure}[tbph]
\includegraphics[width=0.95\linewidth]{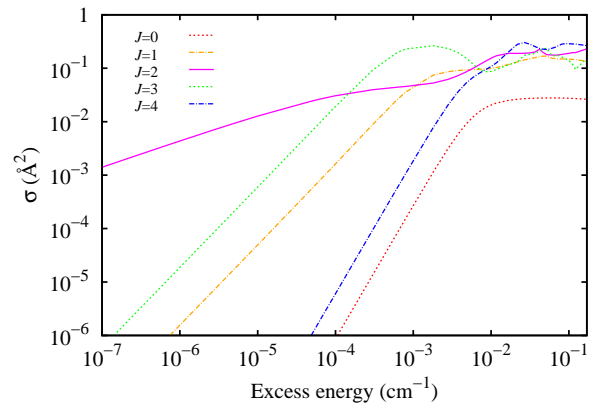}
\caption{Near-threshold behavior of individual partial-wave
contributions to the $11u \to 21l$ inelastic excitation cross
section, from close-coupling calculations, as a function of the
excess energy above the threshold for excitation.}
\label{wigner_open}
\end{figure}

Fig.\ \ref{wigner_open} shows the threshold behavior in more
detail, for the contributions to the cross section
$\sigma_{11u\to 21l}$ obtained from CC calculations for
different partial waves $J$ with the full basis set. CC
calculations are feasible in this case because only low partial
waves contribute just above threshold. At very low energy, each
partial-wave contribution follows the Wigner threshold law
\cite{Wigner:1948} and is proportional to $(E_{\rm coll}-E_{\rm
thresh})^{L_{\rm final}+1/2}$, where $E_{\rm coll}$ and $E_{\rm
thresh}$ are the collision energy and the threshold energy
respectively. $L$ takes the lowest allowed value for the
threshold level: for $j_{\rm final}=2$, $L=0$ for $J=2$, $L=1$
for $J=1$ and 3 and $L=2$ for $J=0$ and 4. For any final $j$
there is always one total angular momentum ($J=j$) and parity
combination that allows $L=0$ in the outgoing channel, and this
dominates the integral cross section for excess energies below
$10^{-4}$ cm$^{-1}$. Above this, however, additional partial
waves contribute and simple $(E_{\rm coll}-E_{\rm
thresh})^{1/2}$ behavior is not expected for the integral cross
section.

For Rb--NH$_3$, the $11u \to 11l $ transition is much stronger
than any other inelastic transition for the whole range of
collision energies considered here. For Rb--ND$_3$,
$\sigma_{11u\to 21l}$ becomes comparable to $\sigma_{11u\to
11l}$ for kinetic energies around 100 cm$^{-1}$. The resonances
in individual state-to-state cross sections are in some cases
quite strong compared to their backgrounds, especially closely
above the thresholds. However, since their magnitude is much
smaller than $\sigma_{11u \to 11l}$, these resonances are not
visible in the total inelastic cross section.

There are clear propensity rules that apply to the cross
sections, as observed previously for He--NH$_3$ \cite{Oka:1968,
Green:1976, Green:1980, Davis:1978} and Ar--NH$_3$
\cite{vanderSanden:1992, vanderSanden:1995} collisions. The
general effect has been discussed by Alexander
\cite{Alexander:1982}. For $k$-conserving collisions,
inversion-changing transitions are preferred when $\Delta j$ is
odd and inversion-conserving transitions are preferred when
$\Delta j$ is even. Thus $\sigma_{11u \to 21l}$ is larger than
$\sigma_{11u \to 21u}$, since the $11u$ and $21l$ channels are
directly coupled by $V_{10}$ potential term. Conversely, for
collisions with $\Delta k=1$, which are driven by the $V_{33}$
potential term, inversion-conserving transitions are preferred
when $\Delta j$ is odd, whereas inversion-changing transitions
are preferred when $\Delta j$ is even. Thus $\sigma_{11u \to
22u}$ is much larger than $\sigma_{11u \to 22l}$. It is
interesting that these propensity rules survive in a system as
strongly coupled as Rb--NH$_3$.

\subsection{Analysis of scattering resonances}

\begin{figure}[tbph]
\includegraphics[width=0.95\linewidth]{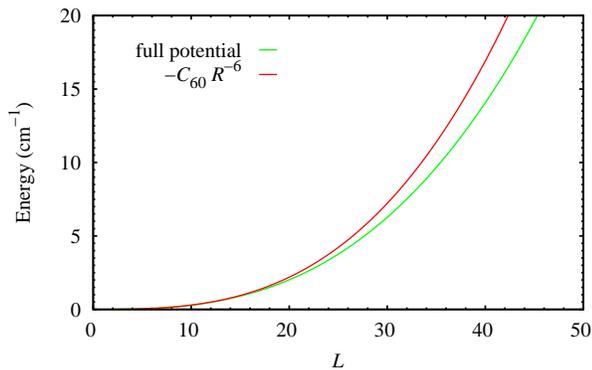}
\caption{Heights of the centrifugal barrier as a function of $L$
predicted by long-range formula (\ref{eq:VL}) and obtained from the
full potential.  } \label{barriers}
\end{figure}

There are in principle two types of resonance that might be
seen in low-energy collisions. Shape resonances correspond to
quasibound states that are confined behind a centrifugal
barrier, whereas Feshbach resonances correspond to quasibound
states that reside principally in channels corresponding to
internally excited states. The height of the centrifugal
barrier for each $L$ is given approximately by
\cite{Hutson:IRPC:2007}
\begin{equation}
V^L_{\rm max} = \left[\frac{\hbar^2 L(L+1)}{\mu}\right]^{3/2}
(54C_6)^{-1/2}. \label{eq:VL}
\end{equation}
This function is compared with the actual barrier heights
obtained from the adiabatic bender curves for Rb--ND$_3$ in
Fig.\ \ref{barriers}; it may be seen that the long-range
formula (\ref{eq:VL}) is accurate for low $L$ but overestimates
the barrier height by about 10\% by $L=30$. However, since
Rb--ND$_3$ has a strongly attractive long-range potential
($C_{60}=523\ E_h a_0^6$) and large reduced mass, the heights
of the centrifugal barriers are much smaller than in lighter
systems. As will be seen below, most of the resonances observed
in the present work are Feshbach resonances.

\begin{figure}[tbph]
\includegraphics[width=0.95\linewidth]{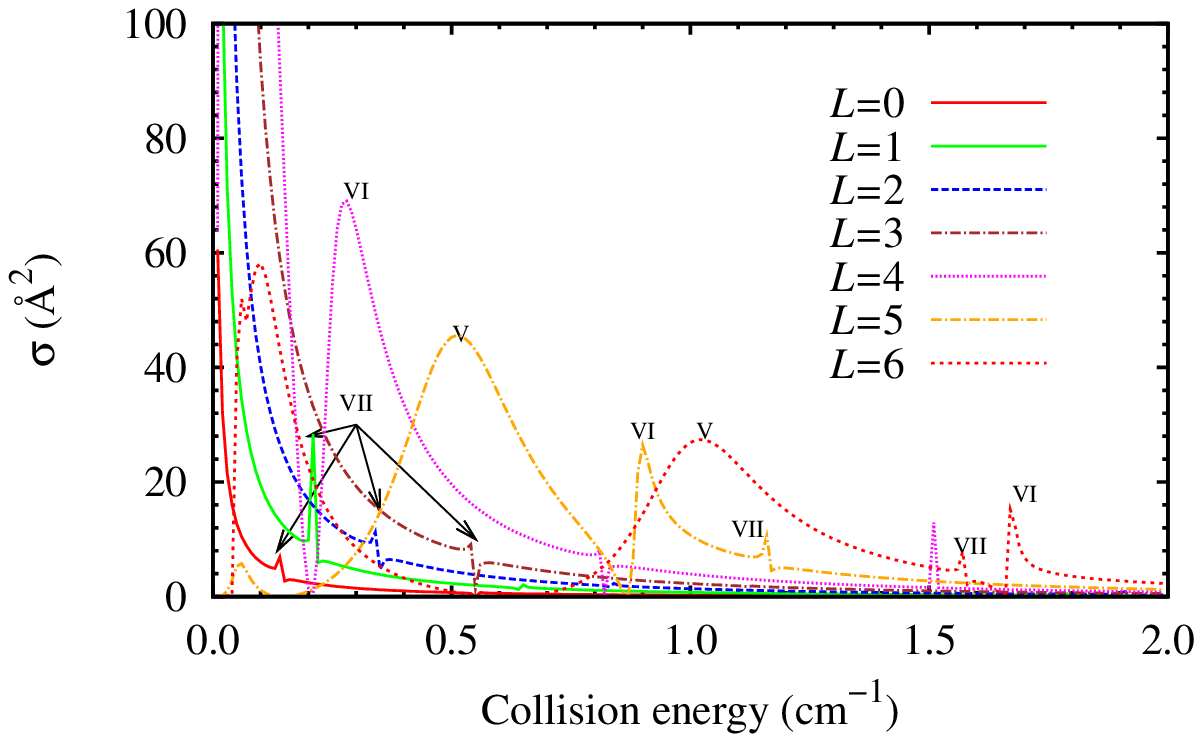}
\includegraphics[width=0.95\linewidth]{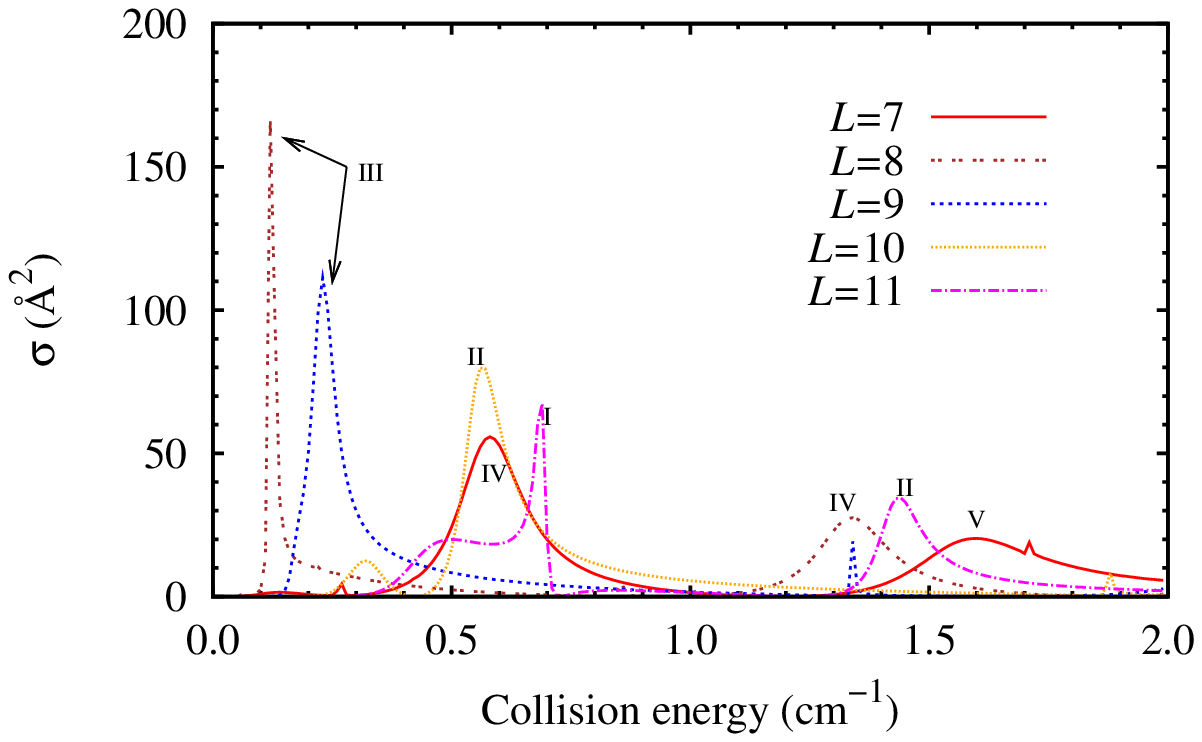}
\includegraphics[width=0.95\linewidth]{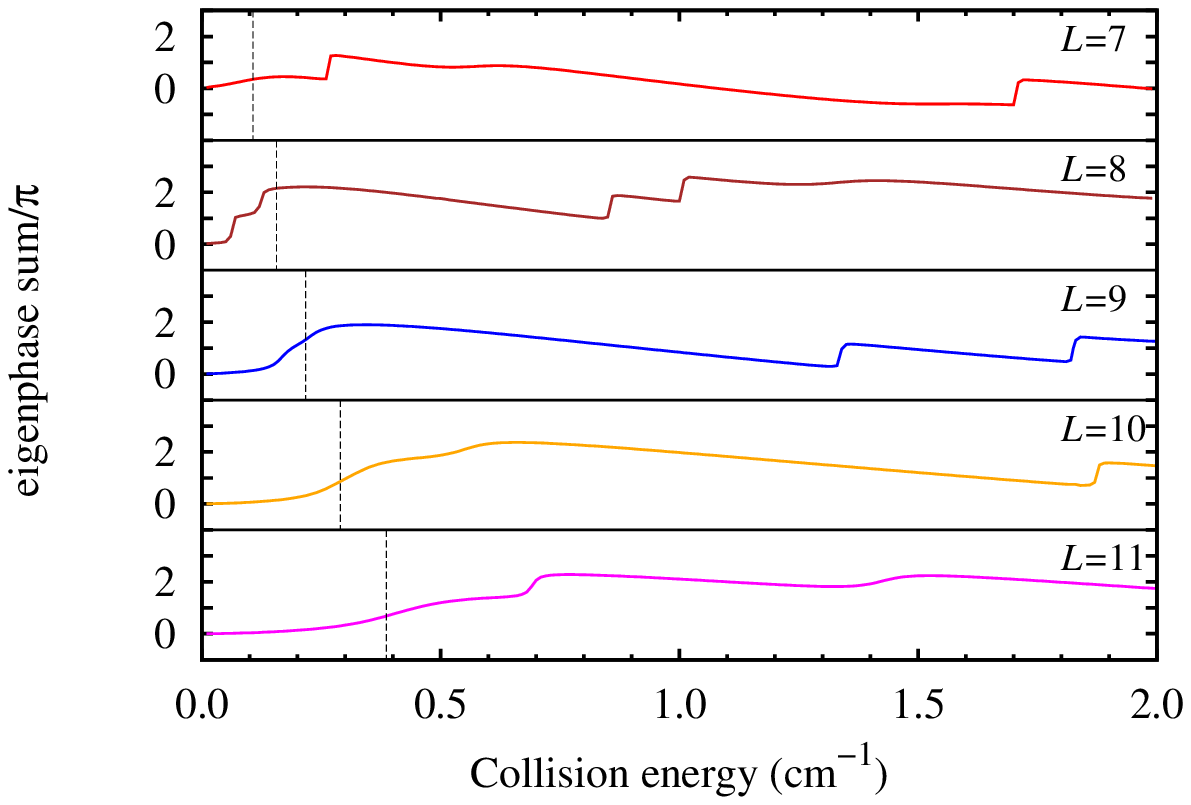}
\caption{Contributions from partial waves $L=0\ldots 6$ (top
panel) and $L=7\ldots 11$ (center panel) to the inelastic $11u
\to 11l$ cross section for Rb--ND$_3$. The resonances are
labeled by Roman numerals corresponding to the series of bound
states shown in Fig.\ {\ref{Bnd}}. The bottom panel shows
eigenphase sums for $L=7\ldots 11$. The dashed vertical lines
in the bottom panel indicate the positions of centrifugal
barriers.}\label{Jexp}
\end{figure}

\begin{figure}[tbph]
\includegraphics[width=0.95\linewidth]{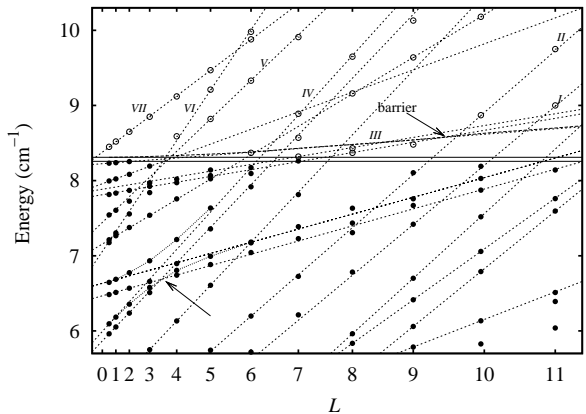}
\caption{Near-threshold bound states of Rb--ND$_3$ (full
circles) for helicity $K=1$ as a function of $L(L+1)$. Open
circles indicate the positions of the strongest resonances in
Fig.\ \ref{Jexp}. The arrow points out an example of an avoided
crossing between bound states. The horizontal solid lines at
8.257 and 8.310 cm$^{-1}$ show the $11l$ and $11u$ thresholds,
respectively. The dashed line shows the position of centrifugal
barrier. } \label{Bnd}
\end{figure}

The origin of the resonances can be understood if we study the
contribution to the inelastic cross sections from individual
terms in the partial-wave expansion. The top two panels of
Fig.\ \ref{Jexp} show the $11u \to 11l$ partial cross sections
for $L=0\ldots 6$ and $L=7\ldots 11$ for Rb--ND$_3$ collision
energies up to 2 cm$^{-1}$, where there are very strong
resonances. The bottom panel shows the S-matrix eigenphase sums
\cite{Ashton:1983} for $L=7\ldots 11$, which show a sharp rise
through $\pi$ as the energy passes through a resonance.

In general each partial inelastic cross section shows a
non-resonant peak near the corresponding barrier maximum and
then dies off at higher energies. The non-resonant peak is
quite sharp for low $L$ but broadens as $L$ increases. In
particular, the peaks below $E_{\rm coll}=0.1$ cm$^{-1}$ for
$L<6$ are non-resonant. Superimposed on the non-resonant
background are peaks and troughs due to resonances. Below the
barrier maximum, both shape and Feshbach resonances may occur;
however peaks due to Feshbach resonances are suppressed when
they are below barriers and it is usually shape resonances that
give large peaks in this region.

Above the barrier maximum for each $L$, Langevin capture theory
would predict a partial inelastic cross section
\begin{equation}
\sigma_{\rm inel}^{\rm capture}(L)=
\frac{\hbar^2\pi L(L+1)}{2\mu E_{\rm coll}}.
\end{equation}
Except at resonances and occasionally near the barrier maximum,
the partial cross sections are generally much smaller than
this. It is thus evident that low-energy inelastic scattering
in Rb--ND$_3$ is {\em dominated} by resonant effects.

We also carried out bound-state calculations close to the
$j=1,k=1$ thresholds to help identify the states responsible
for the resonances. The bound-state calculations used the
coupled-states approximation with the same basis set as for the
collision calculations. The helicity was restricted to $K=1$
since this is the only value that contributes to the $11u \to
11l$ cross section. The resulting bound states are shown in
Fig.\ \ref{Bnd}. Since the rotational energy of the bound and
quasibound states of the Rb--ND$_3$ complex is approximately
proportional to $L(L+1)$, it is convenient to plot the energies
as a function of this quantity. This allows rotational
progressions to be identified as nearly straight lines, and
rotational constants can be extracted from the slopes of the
lines. However, if there are two or more bound states for the
same $L$ with energies very close together, they can mix and
avoid one another; an example of this is shown by the dotted
lines in Fig.\ \ref{Bnd}.

Most of the peaks in Fig.\ \ref{Jexp} can be identified with
series of bound states. For example, the cross sections for
$L=11$ shows three strong peaks. The first of these, at about
0.5 cm$^{-1}$, is a nonresonant peak associated with the
barrier maximum. At higher energies are two resonant peaks,
labeled as {\sc I}, and {\sc II} in Fig.\ \ref{Jexp}. They are
associated with characteristic features in the eigenphase sum
and can be identified as Feshbach resonances associated with
the series of bound states along the dashed lines {\sc I} and
{\sc II} in Fig.\ \ref{Bnd}; series {\sc II} also explains the
strong peak in the cross section for $L=10$ near 0.6 cm$^{-1}$.
Lines {\sc I} and {\sc II} both have a large slope,
corresponding to a rotational constant $B\approx 0.04$
cm$^{-1}$ and an effective distance between Rb and ND$_3$
around $R=9\ a_0$. It is also not difficult to assign the
resonances labeled by {\sc IV, V, VI} and {\sc VII} to the
appropriate series of bound states; all these are Feshbach
resonances that arise from bound states with small effective
intermolecular distances $R<10\ a_0$. For all these bound
states, inspection of the wavefunction near the Van der Waals
minimum confirms that they are dominated by rotationally
excited basis functions with $j>1$.

The levels along the line {\sc III} in Fig.\ \ref{Bnd}, which
give strong resonances for $L=8$ and 9, are associated with
quasibound states of significantly smaller rotational constant,
$B\approx 0.008$ cm$^{-1}$, and an effective intermolecular
distance around 20 $a_0$. The dominant contributions to the
wavefunction in this case are the two $j=1,k=1$ states; the $u$
and $l$ states contribute almost equally and with opposite
signs, corresponding to a single umbrella state of
non-inverting ND$_3$. This suggests that these are shape
resonances, and indeed for $L=8$ and 9 they lie below the
energy of the centrifugal barrier, shown as a dashed curve in
Fig.\ \ref{Bnd}. For $L>9$ the extrapolated energy of line {\sc
III} is above the barrier maximum and no well-defined shape
resonances occur.

It is important to appreciate that the absolute energies of the
resonances and bound states close to threshold are strongly
dependent on the potential energy surface. Thus, the studies
reported in this section should be treated as a paradigm for
understanding the resonant behavior of cross sections, rather
than as predictions of the actual energies at which resonances
will appear for Rb--ND$_3$.

\subsection{Prospects for sympathetic cooling}

As described in the Introduction, there is great interest in
the possibility of {\em sympathetic cooling}, in which
molecules are cooled by thermal contact with a gas of
laser-cooled atoms such as Rb. The simplest sympathetic cooling
experiment would overlap an electrostatic trap containing a
sample of cold molecules such as ND$_3$ with a gas of atoms in
a magnetic or magneto-optical trap. It is therefore of
considerable interest to explore collision cross sections for
promising candidate atoms and molecules in the temperature
regime between 1 $\mu$K and 100 mK.

An electrostatic trap works only for molecules in
low-field-seeking states, and for NH$_3$ and ND$_3$ the
low-field-seeking states correlate at low field with the upper
state of the tunneling doublet. It is thus very important to
know whether collisions with ultracold Rb will cause relaxation
to the lower tunneling state, which is a high-field-seeking
state and cannot be trapped electrostatically. More
specifically, we need to know whether the cross section for the
relaxation from $11u$ to $11l$ state is sufficiently small
compared to the elastic cross section.

\begin{figure}[tbph]
\includegraphics[width=0.95\linewidth]{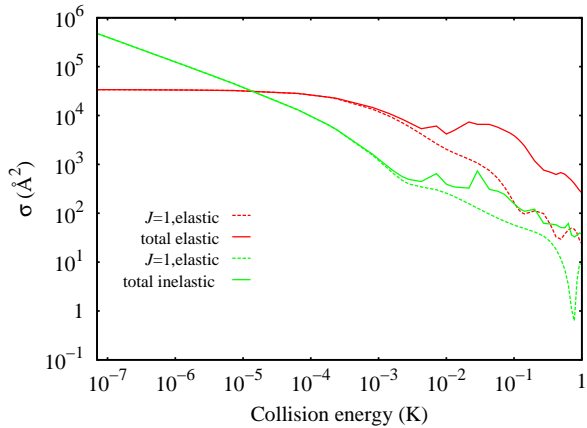}
\caption{Ultracold limit of the elastic $11u \to 11u$ and inelastic
$11u \to 11l$ cross sections and the contribution from $s$-wave
scattering ($J=1$), from close-coupling calculations.}
\label{ultracold}
\end{figure}

To explore this we have performed close-coupling calculations for
very low energies using the same rotational basis set as for the CS
calculations described above. In Fig.\ \ref{ultracold} we show the
elastic and total inelastic integral cross sections calculated for
partial waves $J=0$ to 8, which gives reasonable convergence for
collision energies up to 100 mK. The contribution to the total cross
sections from $s$-wave scattering ($J=1$ partial wave) is dominant
in the $\mu$K regime. The Wigner threshold behavior for $s$-wave
scattering is followed for kinetic energies up to about 100$\mu$K.
Elastic cross sections are larger than inelastic cross sections for
collision energies above 100 $\mu$K, but never by much more than a
factor of 10. For sympathetic cooling, the commonly stated rule of
thumb is that the ratio of elastic to inelastic cross sections
should be at least 100. Thus it appears that the inelastic cross
sections are too large to allow ND$_3$ molecules in
low-field-seeking states to be cooled to sub-mK temperatures by
collision with ultracold Rb atoms.

There remains the possibility of using sympathetic cooling for
molecules in high-field-seeking states, which can be confined
using an alternating current trap \cite{vanVeldhoven:2005,
Bethlem:acNH3:2006}. Confinement of Rb atoms in such a trap has
also been demonstrated \cite{Schlunk:2007}, but the AC
frequencies required are quite different in the two cases. It
is therefore important to know whether sympathetic cooling of
ND$_3$ molecules in high-field-seeking states might be feasible
using magnetically trapped Rb atoms, which are themselves in
low-field-seeking states that are not the ground state in the
applied magnetic field. In previous work on Rb--OH collisions
\cite{Lara:PRL:2006, Lara:PRA:2007}, we found large cross
sections for low-energy collisions that changed the hyperfine
state of Rb. We initially anticipated \cite{Zuchowski:NH3:2008}
that this would occur for Rb--ND$_3$ collisions as well.
However, the more thorough analysis of the collision
Hamiltonian in section II.B above suggests that for molecules
in closed-shell singlet states the atomic spins are likely to
be almost unaffected by collisions. We therefore now consider
that sympathetic cooling of high-field-seeking states of ND$_3$
(or NH$_3$) by magnetically trapped Rb (or another laser-cooled
atomic gas) has a good prospect of success.

\subsection{Sensitivity to the interaction potential}

Collision calculations at ultralow kinetic energies are very
sensitive to details of the potential used in the calculations.
Quantitative theoretical predictions of parameters related to
scattering cross sections, such as the highest bound states or
scattering lengths, are possible only for the simplest,
lightest systems such as metastable helium
\cite{Przybytek:2005,Dickinson:2004} or hydrogen-lithium
mixtures \cite{Gadea:2001}. The calculated Rb--NH$_3$
interaction potential we use in this paper suffers from many
uncertainties resulting from incompleteness of the electronic
basis set, the approximate treatment of electronic correlation,
relativistic effects, etc., which limit its accuracy to at best
a few percent.

\begin{figure}[tbph]
\includegraphics[width=0.95\linewidth]{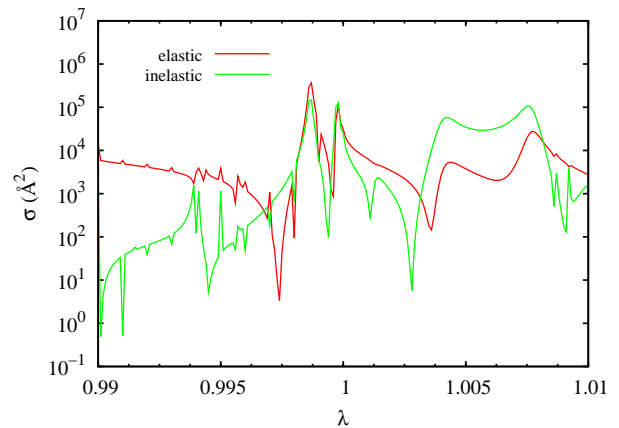}
\caption{Sensitivity of the elastic $11u \to 11u$ and inelastic
$11u \to 11l$ cross sections for Rb--ND$_3$ at a collision
energy of 10 $\mu$K to a scaling factor $\lambda$ applied to
the interaction potential.} \label{scalingparam}
\end{figure}

To explore this sensitivity, we introduce an additional scaling
factor $\lambda$ into the definition of the interaction potential.
We varied the scaling factor between 0.90 and 1.05 and explored the
variation of the $s$-wave elastic $\sigma_{11u \to 11u}$ and
inelastic $\sigma_{11u \to 11l}$ cross sections (using
close-coupling calculations) for a collision energy of 10 $\mu$K.
The results for a representative slice of the range of $\lambda$
studied are shown in Fig.\ \ref{scalingparam}. The elastic and
inelastic cross sections vary dramatically as bound states cross
threshold and appear as scattering resonances. In the range of
$\lambda$ explored, the elastic and inelastic cross sections pass
through more than ten resonances. There are some values of
$\lambda$, such as near $\lambda=0.99$, where the inelastic cross
section $\sigma_{11u \to 11l}$ is significantly suppressed by the
presence of a strong resonance, but for most other values of the
scaling parameter the elastic-to-inelastic ratio ranges lies between
10 and 0.1. Although it is possible that the real potential might
produce inelastic cross sections low enough to allow sympathetic
cooling for low-field-seeking states, it is quite unlikely.

\section{Conclusions}

We have studied collisions between Rb atoms and NH$_3$/ND$_3$
molecules, motivated by recent progress
\cite{Gilijamse:2006,Sawyer:2008} in collision experiments
involving velocity-controlled and Stark-decelerated beams of
molecules and by interest in the possibility of sympathetic
cooling. We focused principally on NH$_3$ or ND$_3$ molecules
initially in the upper component of the inversion doublet for
$j=1$, $k=1$. This level correlates with the low-field-seeking
state in an electric field, which can be slowed by Stark
deceleration and trapped in an electrostatic trap. We
considered collision energies between 0 and 100 cm$^{-1}$.
Using the coupled-states approximation, we calculated the
elastic cross sections and state-to-state inelastic cross
sections from the low-field-seeking $j=1,k=1$ state to other
rotation-inversion levels.

The inelastic cross sections are smaller than expected for such a
strongly coupled system, but are still only about a factor of 10
smaller than the elastic cross section over most of the energy range
considered. Both the elastic and inelastic cross sections show dense
structure due to scattering resonances. The resonances in the
elastic cross sections are diffuse and rather weak compared to the
background. The total inelastic cross sections have much stronger
resonances compared to their background, especially at collision
energies below about 20 cm$^{-1}$ and one can consider the inelastic
scattering as {\em mostly } resonant in nature. These resonances are
washed out for larger collision energies. For energies below 90
cm$^{-1}$ the $11u \to 11l$ inelastic cross section makes the
largest contribution to the total inelastic cross section.
Transitions to $j=2$ and 3 do not change the total inelastic cross
section significantly as the new channels become energetically
accessible at higher collision energies.

We have considered the origin of the scattering resonances,
using calculations of the bound states of Rb--ND$_3$ near
threshold. Since the long-range attraction between Rb and
ND$_3$ is very strong and the reduced mass is fairly large, the
resulting centrifugal barriers are much smaller than in lighter
systems. Most of the resonances appearing in the cross sections
arise from quasibound states in which Rb interacts with
rotationally excited NH$_3$, and can be classified as Feshbach
resonances.

Finally we studied the ultracold limit of the elastic and
inelastic cross sections, in order to test whether
low-field-seeking molecules could be cooled to microKelvin
temperatures by contact with an ultracold gas of Rb atoms.
Hyperfine effects were neglected. We found that inelastic
collisions are strong, and the elastic/inelastic ratio is
unlikely to be sufficient to achieve sympathetic cooling of
ND$_3$ molecules in the $11u$ state. However, there is a good
prospect that sympathetic cooling will be possible for
high-field-seeking states of ND$_3$ or NH$_3$, even if the
atoms used as a coolant are in low-field-seeking states.

\section*{Acknowledgments}
The authors are grateful to Heather Lewandowski for sparking their
interest in Rb--NH$_3$ collisions and to Ruth LeSueur for
discussions on the effect of the ammonia inversion on the
collisions. The authors are grateful to EPSRC for funding of the
collaborative project CoPoMol under the ESF EUROCORES programme
EuroQUAM.

\end{document}